\RequirePackage{ifpdf}
\ifpdf % We are running pdfTeX in pdf mode
\documentclass[pdftex]{sigma}
\else
\documentclass{sigma}

\fi

\begin{document}

\numberwithin{equation}{section}

\allowdisplaybreaks

\renewcommand{\PaperNumber}{046}

\FirstPageHeading

\renewcommand{\thefootnote}{$\star$}

\ShortArticleName{Qualitative Analysis of the Classical and Quantum Manakov Top}
\ArticleName{Qualitative Analysis of the Classical \\ and Quantum Manakov
Top\footnote{This paper is a contribution to the Vadim Kuznetsov
Memorial Issue `Integrable Systems and Related Topics'. The full
collection is available at
\href{http://www.emis.de/journals/SIGMA/kuznetsov.html}{http://www.emis.de/journals/SIGMA/kuznetsov.html}}}

\Author{Evguenii SINITSYN~$^\dag$ and Boris ZHILINSKII~$^\ddag$}

\AuthorNameForHeading{E. Sinitsyn and B. Zhilinskii}

\Address{$^\dag$~Physics Department, Tomsk State University, 634050 Tomsk, Russia}

\EmailD{\href{mailto:evgsin@mail.ru}{evgsin@mail.ru}}

\Address{$^\ddag$~Universit\'e du Littoral, UMR du CNRS 8101, 59140 Dunkerque, France}
\EmailD{\href{mailto:zhilin@univ-littoral.fr}{zhilin@univ-littoral.fr}}

\URLaddressD{\url{http://purple.univ-littoral.fr/~boris/}}

\ArticleDates{Received 20 October, 2006, in f\/inal form 19 January, 2007; Published online March 13, 2007}

\Abstract{Qualitative features of the Manakov top are discussed for the
classical and quantum versions of the problem. Energy-momentum
diagram for this integrable classical problem and quantum joint
spectrum of two commuting observables for associated quantum problem
are analyzed. It is demonstrated that the evolution of
the specially chosen quantum cell through the joint quantum spectrum
can be def\/ined for paths which cross singular strata. The corresponding
quantum monodromy transformation is introduced.}

\Keywords{Manakov top; energy-momentum diagram; monodromy}

\Classification{37J15; 81V55}

\begin{flushright}
\textit{Dedicated to the memory of Vadim Kuznetsov}
\end{flushright}

\renewcommand{\thefootnote}{\arabic{footnote}}
\setcounter{footnote}{0}

\section{Personal introduction}
%This article is dedicated to the memory of Vadim Kuznetsov.

It was about six years ago that I (B.Z.) met for the f\/irst time Vadim
Kuznetsov during one of the
`Geometric Mechanics' conferences in Warwick University, U.K.
I cannot say that
we have found immediately mutual interest in our  research works
in spite of the fact that our problems were rather related.
At that time I tried to understand better the manifestation of
classical Hamiltonian monodromy in corresponding quantum problems
and looked for dif\/ferent simple classical integrable models with
monodromy which could be of interest for physical, mainly molecular,
applications. Vadim worked on much more formal mathematical aspects of
integrable models related to almost unknown for me special functions.
I tried to convince him that from the point of view of physical
applications the most important task is to
understand qualitative features of integrable models using some
simple geometric tools like classif\/ication of defects of regular lattices
formed by joint spectrum of several commuting observables. Vadim
insisted on special functions, complex analysis, Lie algebras etc.
Nevertheless, we have found many points of common
interest.  Soon after,
Vadim visited  Dunkerque and  we have tried to f\/ind some concrete
problem, where we could demonstrate clearly what each of us means by
understanding the solution. In fact, such problem was found quickly.
It was the Manakov top.  Vadim was interested in Manakov top
because of its relation with XYZ Gaudin magnet. He published a short paper
together with I. Komarov on this subject in 1991~\cite{KomKuz91}.
For me the model
problem like Manakov top represented certain interest because it
is naturally related to molecular models constructed,
for example,  by coupled angular momenta or by angular momentum and
Runge--Lenz vectors for hydrogen atom.  In spite of many dif\/ferent
applications and possible generalizations the initial idea was just to
study one concrete simple example which nevertheless keeps all the
important qualitative features in classical and quantum cases
and to show what qualitative aspects of solution seem to
be of primary importance from physical or chemical point of view.
Unfortunately, this paper is written when Vadim is gone away and
we will not hear his criticism and ref\/lections about molecular
physicist point of view on the integrable quantum Manakov top.

\section{The model}
In this article we will study one concrete example of the Euler--Manakov
top having a complete set of quadratic integrals of motion. We analyze
both classical and quantum versions and make below no dif\/ference
in notation between
quantum operators of angular momenta and their classical counterparts.
The integrable Manakov top \cite{Manakov} and its various
generalizations were studied on dif\/ferent occasions but mainly
within classical mechanics \cite{Adler,Audin,BolFomBook,%
DavDulBol,DavDul,Oshemkov,Perelom}.

We def\/ine the Manakov top in accordance with \cite{Kalnins,KomKuz91}
as a system of two
commuting quadratic functions on the ${\rm o}(4)$ generators ($a$ and $b$ being
arbitrary real constants)
\begin{gather}
 X=s_1t_1+\frac{a-b-1}{1-a-b}s_2t_2+\frac{b-a-1}{1-a-b}s_3t_3, \nonumber
 \\
 Y=b(1-a)(s_2^2+t_2^2)+2 b(1-a)\frac{b-a-1}{1-a-b}s_2t_2 \nonumber
+a(1-b)(s_3^2+t_3^2)
\\ \phantom{Y=}
{}+2 a(1-b)\frac{a-b-1}{1-a-b}s_3t_3.
\label{eq_XY}
\end{gather}
Here the generators $s_i$, $t_i$, $(i=1,2,3)$ obey the standard commutation
relations (${\rm o}(4) \simeq {\rm su}_s(2)\otimes {\rm su}_t(2)$):
\begin{gather*}
%\label{eq_relationST}
[s_i,s_j]=i\varepsilon_{ijk}s_k, \qquad [t_i,t_j]=i\varepsilon_{ijk}t_k,
\qquad [s_i,t_j]=0.
\end{gather*}
Two commuting integrals of motion $X,Y$ ($[X,Y]=0$) and two f\/ixed values of
$\bar{S}^2=\sum_i s_i^2$ and
$\bar{T}^2=\sum_i t_i^2$  (or equivalently of two Casimir operators
of ${\rm o}(4)$) f\/ix the system. For further
simplicity we will always choose the normalization
$\bar{S}^2=\bar{T}^2=1$ in classical model and make the
necessary scaling of  quantum angular momentum operators in order to
increase the density of quantum eigenvalues in the case of quantum
calculations. We use as usual in quantum mechanics of angular momentum
the quantum numbers $S$ and $T$ which are related to eigenvalues of
the $S^2=s_1^2+s_2^2+s_3^2$ and $T^2=t_1^2+t_2^2+t_3^3$ operators
as $S(S+1)$ and $T(T+1)$.

The global idea of the present analysis is to compare the qualitative
features of classical energy-momentum (EM) diagram with the qualitative
features of associated joint spectrum of two commuting quantum
observables and to stimulate the discussion about possible
generalization of the
monodromy concept to problems which have several connected components
in the inverse image of the energy-momentum map and admit the presence
of certain codimension one singularities associated with the fusion of
dif\/ferent components.

In the main text below we only discuss the most essential
qualitative aspects of
the classical EM diagram and of the quantum joint spectrum and
formulate several questions about qualitative features which
still remain unclear, at least for the authors, and probably
require more sophisticated qualitative mathematical arguments to
answer. In appendices, we brief\/ly explain some simple tools
which were used to recover the qualitative description of the
studied problem.

\newpage

\section{General overview of the energy-momentum diagram}
For integrable problems the extremely useful geometrical representation
consists in constructing the image of the energy momentum map,
which in classical problems is often named as energy-momentum
diagram, or bifurcation diagram, while in quantum mechanics
it is the geometrical representation of the joint spectrum of
commuting quantum observables.

\begin{figure}[t]
\centerline{\includegraphics[width=7.5cm]{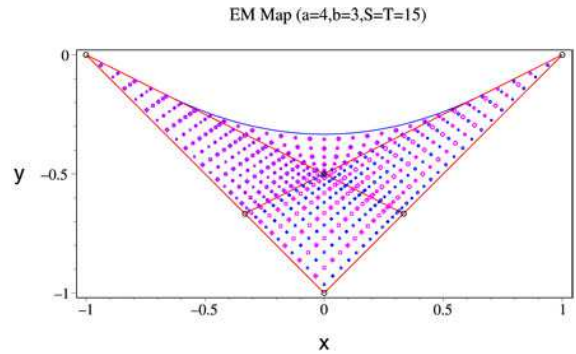}}
\caption{Classical energy-momentum diagram and joint spectrum of two
integrals of motion for the Manakov top model (\protect{\ref{eq_XY}}) with
$a=4$, $b=3$, $S=T=15$.}
  \label{fig_EMtot15}
\end{figure}

For Manakov top problem (\ref{eq_XY}) with one concrete choice of parameters
$(a=4$, $b=3$, $S=T=15)$ the classical energy-momentum diagram
is represented in Fig.~\ref{fig_EMtot15} together with the
joint spectrum of two commuting quantum operators. We have chosen the
values of parameters to produce the most symmetric form of the image of
EM map and to have a suf\/f\/icient number of quantum states to see the
characteristic pattern in each qualitatively dif\/ferent part of the
diagram.

We know that in general for completely
integrable problems with two degrees of freedom the image
of the EM map def\/ines the foliation of the classical phase space
(which is the $S^2\times S^2$ for the
Manakov top for the given choice of the
Casimir values) into common levels of two integrals of motion $X$, $Y$
which are in involution by construction. All possible values of
the EM map  f\/ill in the $X$, $Y$ plane (see Fig.~\ref{fig_EMtot15})
the `curved triangular region'
which consists of regular and singular values. Singular values belong
to special lines and points. Regular values f\/ill 2-D-regions.
Any regular value of the EM map for problem with two degrees of freedom
has as its inverse image one or several two-dimensional tori.
Singular values can have dif\/ferent types of f\/ibers as inverse images.
All these f\/ibers can be generally described as singular tori. Due to that
we characterize qualitatively the whole foliation
as `singular f\/ibration'.

Before starting to discuss qualitative features of the concrete
problem we need to verify that the chosen problem is generic or
structurally stable. In other words we need to verify that any admissible small
modif\/ication of parameters does not change qualitative features
of classical energy momentum diagram and of joint quantum spectrum.

The f\/irst initial simplest qualitative characterization of both the
EM map image and
the foliation  is the splitting of the whole image of
the EM map into connected regions of regular values. Each region
is characterized by the number of connected components (tori)
in the inverse image of  each regular value.

\begin{figure}[t]
\centerline{\includegraphics[angle=270,width=6cm]{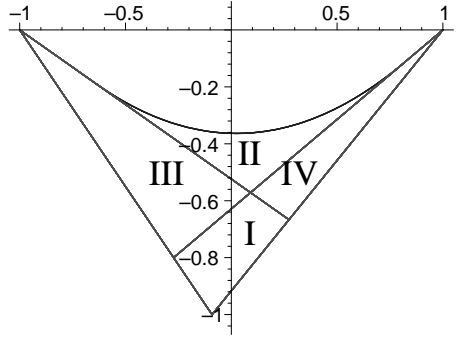}}
\caption{Base of integrable f\/ibration of Manakov top (bifurcation diagram).}
  \label{fig_EMgencase}
\end{figure}

Fig.~\ref{fig_EMgencase} shows what happens when we change slightly
parameters $a$, $b$ of the model (\ref{eq_XY}). We keep, naturally, under such
deformation the integrability of the problem and the $S=T$
condition. The image
of the EM map becomes less symmetric but it  keeps the presence
of four regions of regular values and the numbers of connected components in
regular inverse images. There are two  connected components in regions labeled by
I, III, and IV, and four connected components in region II.

\begin{figure}[t]
\centerline{\includegraphics[width=10cm]{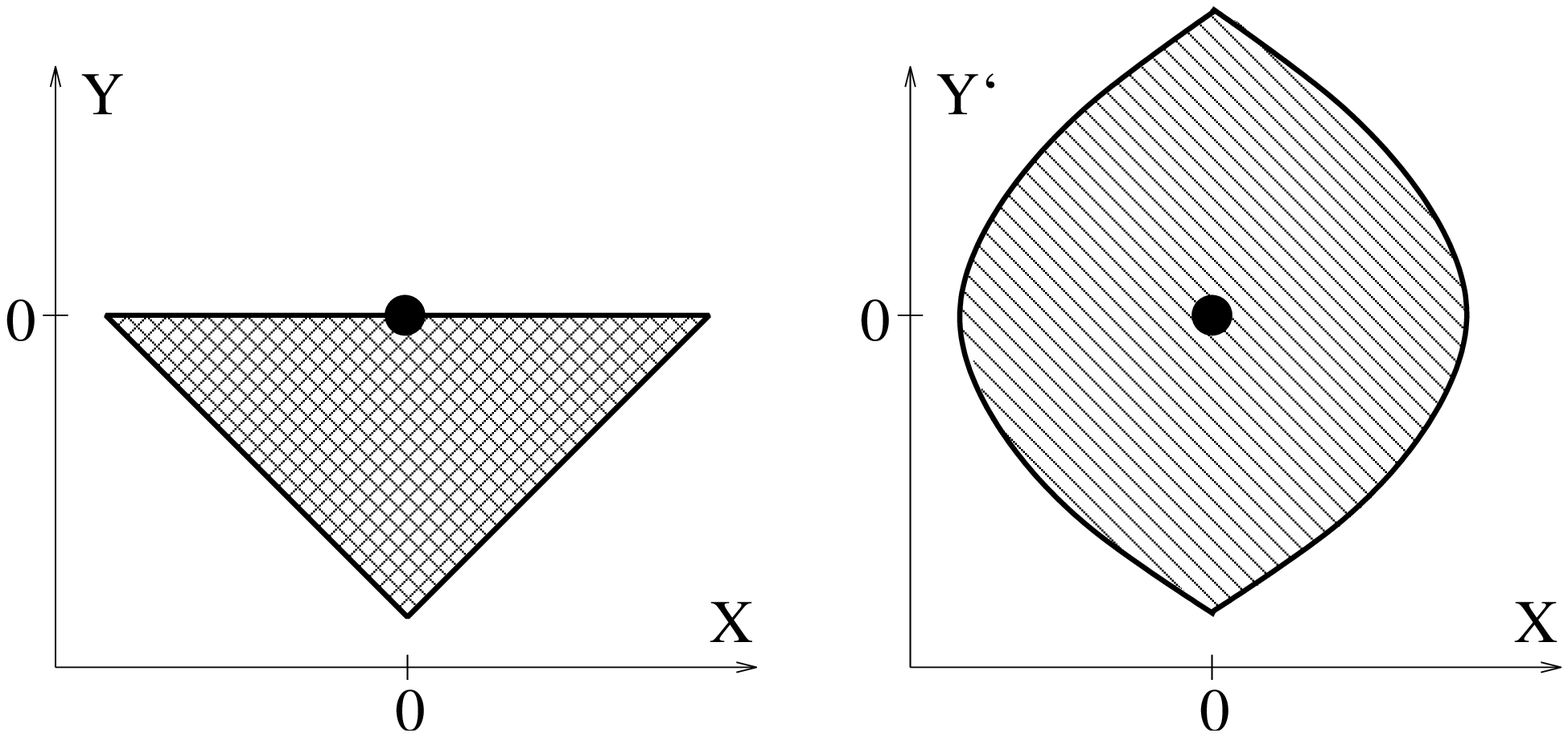}}
  \caption{Left: Base of integrable f\/ibration of Manakov top for a special
limiting case ($a=2$, $b=1$). Right: Unfolded bifurcation diagram for
($a=2$, $b=1$) Manakov top with $Y^\prime=\sqrt{-Y}$.}
  \label{fig_Limcase}
\end{figure}

As soon as small variation of parameter values $a$, $b$ does not modify the number of
regions on the bifurcation diagram the chosen values can be considered as
generic. At the same time large variation of the same parameters can
lead to qualitatively dif\/ferent bifurcation diagrams. Various possibility
are discussed in the appendix. We mention here only one important
limiting case corresponding to $(a=2$, $b=1)$. This limiting case is
extremely simple because of particular form of two commuting
operators $X$, $Y$:
\begin{gather}
\label{Limit}
X = s_1t_1+s_3t_3,\qquad
Y = - \frac{(s_2+t_2)^2}4.
\end{gather}
The corresponding bifurcation diagram is shown in Fig.~\ref{fig_Limcase},
left. It follows from generic bifurcation diagram (Fig.~\ref{fig_EMgencase})
by shrinking
regions II, III, and IV to zero. In this limiting case we have only
one region with each internal point corresponding to two connected
components (regular tori) in the inverse image.

By going to $Y^\prime = \sqrt{-Y}$ we easily recover the integrable
f\/ibration with one connected component in the inverse images of any point and
the singularity at $Y^\prime=X=0$ corresponding in the
classical case to doubly pinched torus. The transformation from
$\{X,Y\}$ to $\{X,Y^\prime=\sqrt{-Y}\}$ can be described as `square root
unfolding'. It was analyzed in classical mechanics recently \cite{DavDulBol}.
The corresponding transformation of the bifurcation diagram is shown in
Fig.~\ref{fig_Limcase}.

We make accent in the present paper on the analysis of the quantum problem.
But in order to recover the qualitative features of the quantum joint
spectrum we need to study it together with the classical bifurcation
diagram.

\begin{figure}[ht]
\centerline{\includegraphics[width=8cm]{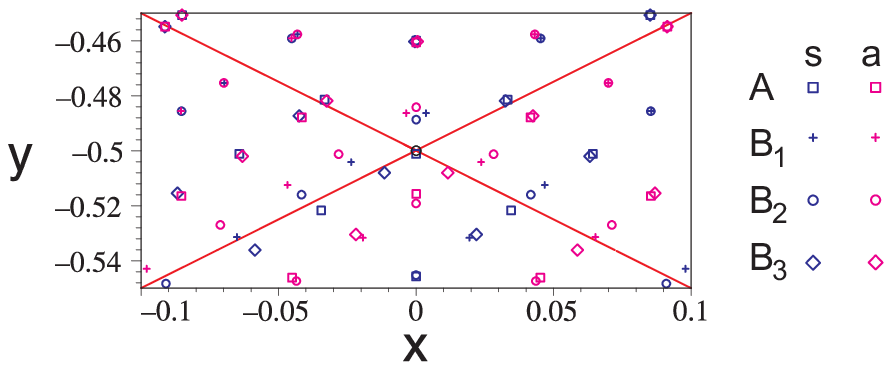}}
  \caption{Central part of the joint spectrum of the Manakov top
 shown in Fig.~\protect{\ref{fig_EMtot15}}.}
  \label{Central}
\end{figure}

The joint spectrum of quantum operators is shown in the same Fig.~\ref{fig_EMtot15} by dif\/ferent symbols depending on the symmetry
of the state. Taking into account the f\/inite symmetry of the
problem (see appendix for details) there are quantum states of
eight dif\/ferent symmetry types. It is practically impossible to distinguish
dif\/ferent symbols on this Fig.~\ref{fig_EMtot15} because of the
overlapping of dif\/ferent symbols due to quasi-degeneracy of eigenvalues.
All representations of the symmetry group are one-dimensional, but the
quasi-degeneracy is almost perfect in the most part of the region (except
neighborhoods of internal singular lines). Thus it is clear that
the most prominent qualitative feature of the joint spectrum is
its cluster structure. The rearrangement of clusters near singular lines
is more clearly seen in Fig.~\ref{Central} which shows in more
details the joint quantum spectrum in the most complicated central part
of the energy momentum diagram.

\begin{figure}[ht]
\centerline{\includegraphics[width=11cm]{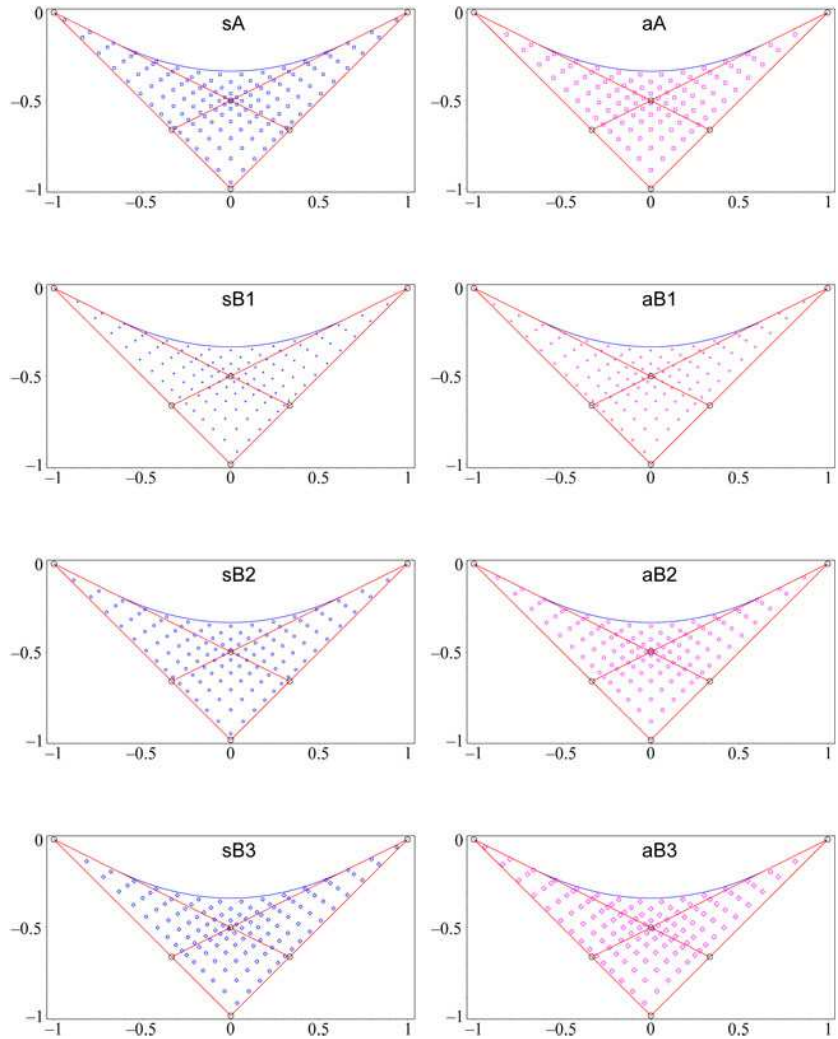}}
 \caption{Joint spectrum  of Manakov top. Only eigenvalues of one
   symmetry type are shown on each sub-f\/igure.}
  \label{fig_sA}
\end{figure}

We can neglect in the f\/irst
approximation the internal structure
of clusters and try to cha\-racterize clusters and to understand their
global arrangement. It is quite easy to see that in regions of
EM diagram with $K$ connected components for each inverse image,
the $K$-fold clusters of quantum eigenvalues should be present.
Thus in regions I, III, IV of the EM diagram the 2-fold clusters are
present, whereas the region II is f\/illed with the 4-fold clusters.
Highly regular pattern formed by the
common eigenvalues is clearly seen in  Fig.~\ref{fig_EMtot15}.
The most part of each of the four regular regions can be regarded as
covered with almost regular lattice of 2-fold or 4-fold clusters.
By a slight
deformation each such lattice can be deformed to a part of an ideal
square lattice.
Rearrangement of clusters takes place near the lines of singular
values of classical EM map and an apparent
non-regularity is concentrated near the singular lines (see Fig.~\ref{Central}).

\looseness=1
In order to understand the origin of regularity and non-regularity we
remind f\/irst that the classical phase space of the Manakov top problem
is compact and the total number of quantum eigenstates is f\/inite
and is determined by quantum numbers $S=T$. The example shown in
Fig.~\ref{fig_EMtot15} corresponds to the choice $S=T=15$.
This means that the
total number of quantum eigenvalues is $31\times31=961$. If we denote
eight dif\/ferent irreducible representations by $A_s$, $A_a$, ${B_i}_s$,
${B_i}_a$, with $i=1,2,3$, the numbers of eigenvalues of each type
of symmetry are: $14\times15/2=105$ for $A_a$,
$15\times16/2=120$ for ${B_i}_s$, ${B_i}_a$, $i=1,2,3$,
and $16\times17/2=136$ for $A_s$.
If now we plot eigenvalues of only one type of the symmetry
on the classical energy-momentum diagram the pattern of quantum
states turns out to be almost regular in the whole region.
Fig.~\ref{fig_sA} demonstrates this fact for all eight symmetry
types. The observation of the regularity of joint spectrum for one
symmetry type is based on numerical results and requires further
independent explanation.

At the same time if we analyze the total joint spectrum
(see Fig.~\ref{fig_EMtot15}) it is clear that in regions
I, III, IV (two triangle regions and rhomb region, see
appendix) the joint spectrum can be
qualitatively described as a regular
doubly degenerate lattice of common eigenvalues.
Region II
(parabola region in the notation of appendix) of the joint
spectrum in a similar way can be qualitatively characterized
as formed by four-fold degenerate regular lattice of common
eigenvalues.

Moreover, pairs of quasi-degenerate eigenvalues form dif\/ferent
reducible representations in dif\/ferent regions.
In region I the degenerate pairs form $A_s+B_{2s}$,
$A_a+B_{2a}$, $B_{1s}+B_{3s}$, and $B_{1a}+B_{3a}$
representations.
Degenerate pairs in region III are
$A_s+B_{3a}$, $B_{1s}+B_{2a}$, $B_{2s}+B_{1a}$, and
$B_{3s}+A_a$.  In region IV we have
$A_s+B_{3s}$, $B_{1s}+B_{2s}$, $A_a+B_{3a}$, and
$B_{1a}+B_{2a}$.
The quadruples of eigenvalues in region II are
$A_s+A_a+B_{3s}+B_{3a}$ and $B_{1s}+B_{1a}+B_{2s}+B_{2a}$.

\looseness=1
It is clear that the lattices def\/ined in dif\/ferent
regions cannot f\/it together along the singular lines
because of dif\/ferent organization of the joint spectrum
in dif\/ferent regions and because of dif\/ferent numbers of
states of various symmetry types. At the same time the regularity
of the joint spectrum for each symmetry type indicates the
possibility of the global regular labeling of states
of each particular symmetry type.
Using such labeling we can try to def\/ine the continuous
evolution of the properly chosen quantum cell through the
quantum joint spectrum along the path which crosses  singular
lines. This is exactly what is needed in order to def\/ine
quantum monodromy using approach based on the propagation of
the elementary quantum cell. Before doing that
we remind shortly in the next section the description of
quantum monodromy and its generalizations
\cite{Grondin,AnnHP,SZhPhysL,MolPhys,ZhilAAM,ZhilTOP} in terms of
the `quantum cell' evolution.

\section{Relation to monodromy and its generalizations}
Qualitative description of joint spectra of
quantum problems associated to some integrable classical
models is based on the simultaneous description of defects
of regular lattices and on af\/f\/ine structure of classical
models.

Presence of singular f\/ibers for classical integrable f\/ibrations leads
to the appearance of defects of the lattice of
joint eigenvalues in corresponding quantum problem.
For an integrable Hamiltonian system with two degrees of freedom
the simplest codimension two singularity which generically
appears in the bifurcation diagram
is the so-called focus-focus singularity.
The asso\-cia\-ted singular f\/iber is a pinched torus. Several times
pinched tori are typically possible under the presence of
symmetry.
Recent rigorous mathematical description of dynamical problems
with such singularities can be found in
\cite{Symin2,Symin1,san,SanAdv}.
More physically based intuitive construction was suggested in
\cite{Grondin,SZhPhysL,ZhilAAM,ZhilTOP}.

Classical monodromy which is due to the presence
of a focus-focus singularity (pinched torus) manifests
itself in the corresponding quantum problem in the most clear and
transparent way as a~transformation of the elementary cell of
the lattice formed by the joint eigenvalues of commuting
quantum observables after its propagation along a closed path
surrounding the singularity.
The evolution of an elementary quantum cell is based on the existence
of local action-angle variables, whereas the nontrivial monodromy
demonstrates the absence of global action-angle variables.

\looseness=1
Although the relation between Hamiltonian monodromy and the absence
of global action-angle variables \cite{BlueBook}
was formulated  in the  well-known
papers by Nekhoroshev \cite{NNaav} and Duistermaat~\cite{Duist80}
about 30 years ago the signif\/icant physical applications of
hamiltonian monodromy in such simple physical systems as
atoms and molecules were found only in the last ten years.
 Without going into details of physical applications we just cite
here several recent papers where important physical applications and
citations to other publications can be found
\cite{NuclPhys,Child,CushSadHatom,AnnHP,MolPhys,AnnPhys,Winnew}.
From the mathematical point of
view the generalization of monodromy can go in two dif\/ferent
directions. One can try to def\/ine the evolution of a quantum cell or
of bases of homology groups for classical integrable f\/ibrations
when the closed path crosses some special singular strata.
On this way the notions of fractional monodromy
\cite{ECSfrac,NSZCR,AnnHP} and of bidromy \cite{MolPhys,AnnPhys}
were introduced. Completely another possibility is to
study higher obstructions to the existence of the action-angle
variables, which are related to the codimension-$K$ singularities
with $K\geq 3$ \cite{Duist80}. We does not touch this aspect here. Instead
we suggest on the example of the Manakov top the generalization
of the monodromy notion to a  larger class of dynamical
systems (integrable f\/ibrations) which admits the presence
of new passable singular strata,
associated with fusion or splitting of several connected
components of the inverse EM image into one.

\section{Quantum monodromy for Manakov top}

In this section we formulate new result about propagation of  quantum cells
along noncontractible paths in the base of integrable singular f\/ibration
def\/ined by  the Manakov top problem. In order to do that we need f\/irst to
def\/ine the path itself. The main problem here is due to:
\begin{enumerate}
\itemsep=0pt
\item[i)] the presence of
two or four connected components of the inverse EM image
in dif\/ferent regular regions of bifurcation diagram;
\item[ii)] the continuation of the path when crossing singular strata.
\end{enumerate}

\begin{figure}[t]
\centerline{\includegraphics[width=9cm]{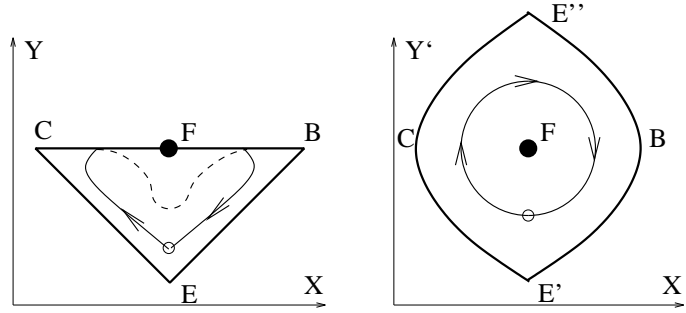}}
\caption{Choice of the closed path on the base space for the
limiting case $a=2$, $b=1$ of the Manakov top. Left: Representation in
$\{X,Y\}$ variables with two components of the inverse image for each
regular point. Solid and dashed lines represent parts of the path which
belong to dif\/ferent leafs of the unfolded diagram.
Right: Representation in $\{X,Y^\prime=\sqrt{-Y}\}$ variables with one
component of the inverse image for any regular point. }
\label{Path}
\end{figure}

\looseness=1
We start with a simple limiting case $a=2$, $b=1$ of the Manakov top problem.
Corresponding classical integrable f\/ibration
is represented by its bifurcation diagram in $\{X,Y\}$
variables in Figs.~\ref{fig_Limcase}, left and~\ref{Path}, left.
This f\/ibration  has two components of the inverse EM image for all regular
values. At the same time the inverse image of each of singular values on
$(CF)$ and $(FB)$ intervals is one regular torus. Such structure gives possibility
to unfold the picture by going from $\{X,Y\}$ to $\{X,Y^\prime=\sqrt{-Y}\}$
variables \cite{DavDulBol}.
The closed path in the unfolded variables $\{X,Y^\prime\}$
(see Fig.~\ref{Path}, right) encircles the isolated singular value
(point $F$) whose inverse image is a doubly pinched torus.
Continuous evolution of classical f\/ibers and of corresponding quantum cell
along such contour should necessarily lead to nontrivial monodromy~\cite{DavDulBol}.
We demonstrate this below on the joint spectrum of this particular example.
For a moment we just stress the representation of the closed path in
the original variables $\{X,Y\}$ in Fig.~\ref{Path}, left.

\begin{figure}[t]
\centerline{\includegraphics[width=7cm]{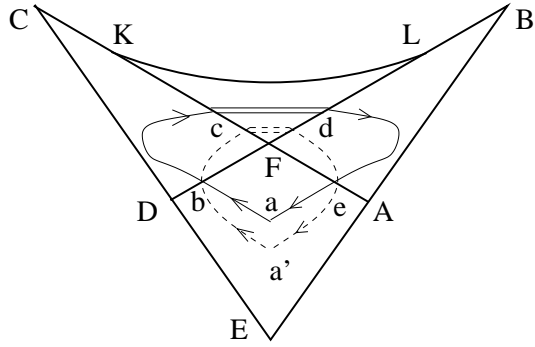}}
\caption{Choice of the generalized closed path on the base space for the
generic Manakov top problem.}
\label{PathG}
\end{figure}

When constructing the closed path on the base of the classical
foliation of the generic Manakov top (see Fig.~\ref{PathG}) we need
to remind that in the internal points of regions $AFDE$, $CDF$, and $AFB$
(regions I, III, IV, in Fig.~\ref{fig_EMgencase} respectively)
the inverse image of the EM map has
two connected components, whereas in the internal points of region $KLF$
(region II in Fig.~\ref{fig_EMgencase})
there are four components. At intervals $DF$ and $FA$ each point has
one connected component in its inverse image. At the same time
at the boundary $LBAEDCK$ there are two components except for
points $A$ and $D$. Intervals $KF$ and $FL$ are associated with
singular f\/ibers corresponding to splitting of each connected component
presented for internal points in regions $AFB$ and $CDF$ into two
regular components in region $KLF$.

We start the path at point $a$ in the `rhomb' region
$AFDE$ (region I), see Fig.~\ref{PathG}. As soon as there
are two connected components in the inverse image, we need to
distinguish them and to precise that the point $a$
belongs to one of these components, say to leaf I$_1$.
Alternatively, we can choose starting point $a^\prime$ at another
leaf I$_2$. Two dif\/ferent paths associated with dif\/ferent components
are represented in Fig.~\ref{PathG} by solid and dash lines.

When going through the point $b\in (DF)$ two components fuse
together and split again into two components of the region III,
which we denote as III$_1$, and III$_2$. Crossing singular stratum~$DF$
needs further analysis. We suppose for a moment that
at point $b$ we can follow the path
continuously from I$_1$ to III$_1$ and in a similar way from I$_2$ to
III$_2$. The intuitive arguments for such continuation will be given
below on the basis of the analysis of the evolution of quantum joint
spectrum in the neighborhood of singular stratum $DF$.

Next essential step is the crossing the singular stratum $KF$.
When crossing this stratum one component splits into two connected
components and one regular torus transforms into two
regular tori. Similar situation was studied recently  by
Sadovskii and Zhilinskii \cite{AnnPhys}, who introduced the notion of
`bidromy' and the associated notion of `bipath' when crossing the
singular stratum corresponding to splitting of one regular component
into two components. In a similar way the path
which is def\/ined in the region
III  and follows the component III$_1$ splits into two-component-path
when entering region IV by crossing singular stratum $KF$.
This transformation is represented in Fig.~\ref{PathG} by
going from single solid line to double solid line after crossing
$KF$. Analogous but independent transformation takes place for
the path def\/ined in region III for the component III$_2$ (dash line in
Fig.~\ref{PathG}).

Further continuation through singular strata $FL$ and $AF$ enables us to
def\/ine the closed path starting at point $a$ on component I$_1$ and
ending at the same point. Again the intuitive justif\/ication of the construction
of such generalized closed path is given below on the basis of the
evolution of `quantum cells' through the joint spectrum of mutually
commuting quantum operators for Manakov top problem.

Obvious dif\/f\/iculty in proper def\/inition of the corresponding classical
(quantum) construction for such generalized
closed path is related to the def\/inition of the connection
between bases of the homology groups (or appropriate subgroups)
of regular f\/ibers when
the path crosses singular strata. In this article we do not want to
discuss this delicate question and leave the description of the
classical problem open for further study. In contrast, we
concentrate on the quantum problem and demonstrate below how
quantum mechanical results allow to introduce the continuous
evolution of the quantum cell through singular strata and to
def\/ine the quantum monodromy for the Manakov top.

We hope that the quantum aspect can stimulate further
analysis of corresponding classical problem which will probably
lead to another generalization of the classical Hamiltonian monodromy
concept in a way similar to appearance of classical fractional Hamiltonian
monodromy stimulated by initial quantum conjectures.

\looseness=1
The key point in the construction of the evolution of the quantum
cell through the joint spectrum is the regularity of the pattern
of common eigenvalues of two commuting operators formed by
eigenvalues with one chosen symmetry type. It is useful to remind here
that in almost all standard problems with Hamiltonian monodromy
one of the integrals is related to continuous symmetry and is a good global action
variable by construction. In such a case splitting the total
set of common eigenvalues into subsets with dif\/ferent symmetry types
leads from 2D-pattern for the total problem to a family of 1D-patterns
for dif\/ferent symmetry types. In order to see monodromy we are
obliged to compare common eigenvalues with dif\/ferent continuous
symmetry (dif\/ferent values of angular momentum, for example).
Fractional monodromy also follows naturally this kind of reasoning.
Looking at two sub-lattices  separately (for problems possessing
half-integer monodromy)
does not allow to see the phenomenon (see detailed discussion in~\cite{AnnHP}). In order  to observe the half-integer monodromy
we are obliged to analyze two (index two) sub-lattices
simultaneously.

The Manakov top problem possesses f\/inite symmetry group which
allows classif\/ication of the common eigenvalues of operators $X$
and $Y$ by eight dif\/ferent irreducible representations. This allows us
to make a choice of elementary quantum cell in the region I as
formed by four dif\/ferent eigenvalues.

Before going to the analysis of the evolution of quantum cell for generic
Manakov top problem we study f\/irst one particular limiting case
$a=2$, $b=1$ (see equations~(\ref{Limit}) and
Figs.~\ref{fig_Limcase}, \ref{Path}) of the Manakov top problem.

%%%%%%%%%%%%%%%%%%%%%%%%%%%%%%%%%%%%%%%%%%%%%%%%%%%%%%%%%%%%%%%%%%
\subsection{Quantum cell evolution for particular limiting case}
The particular case $a=2$, $b=1$ of the Manakov top problem is
especially simple because of a~possibility of a `square root unfolding'
\cite{DavDulBol}  by going to new commuting variables
$X=s_1t_1+s_3t_3$, $Y^\prime=\sqrt{-Y}=(s_2+t_2)/2$.

\begin{figure}[t]
\centerline{\includegraphics[width=6cm]{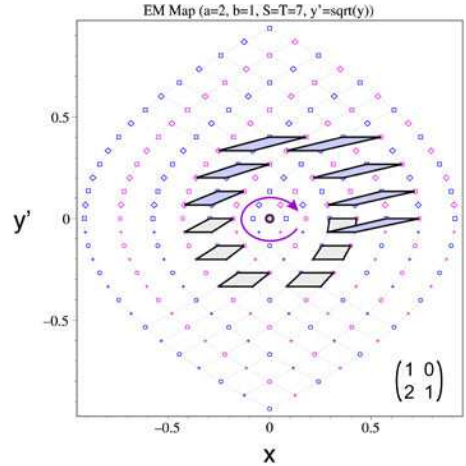}}
\caption{Joint spectrum of commuting quantum operators
$X,Y^\prime = \sqrt{-Y}$ for the limiting case $a=2$, $b=1$ of the Manakov top
together with
the evolution of the elementary quantum cell along a closed path encircling
critical value $X=Y^\prime=0$.}
\label{F:LimCase}
\end{figure}

The evolution of the elementary quantum cell through the
joint spectrum of $X$, $Y^\prime$ becomes  simple because $Y^\prime$
is the generator of a global continuous symmetry, the projection of the
total momentum on axis 2. In classical mechanics $Y^\prime$ can be used
as a global action. Second global action does not exist for this
problem due to presence of an isolated critical value $X=Y^\prime=0$.
Fig.~\ref{F:LimCase} clearly shows that the monodromy matrix
corresponding to the transformation of the elementary cell along a closed
path encircling the critical value in the chosen basis of the joint spectrum
lattice has the form
$\left(\begin{matrix} 1 & 0 \\ 2 & 1 \end{matrix} \right) $.

It is useful to go back to original $\{X,Y\}$ variables and to represent
the evolution of the quantum cell along the same closed path but
in $X$, $Y$ variables as in Fig.~\ref{Path}, left.

%%%%%%%%%%%%%%%%%%%%%%%%%%%%%%%%%%%%%%%%%%%%%%%%%%%%
%%%%%%%%%%%%%%%%%%%%%%%%%%%%%%%%%%%%%%%%%%%%%%%%%%%%%%%%%%%%%%%%%

\begin{figure}[t]
\begin{minipage}{75mm}
\includegraphics[width=7cm]{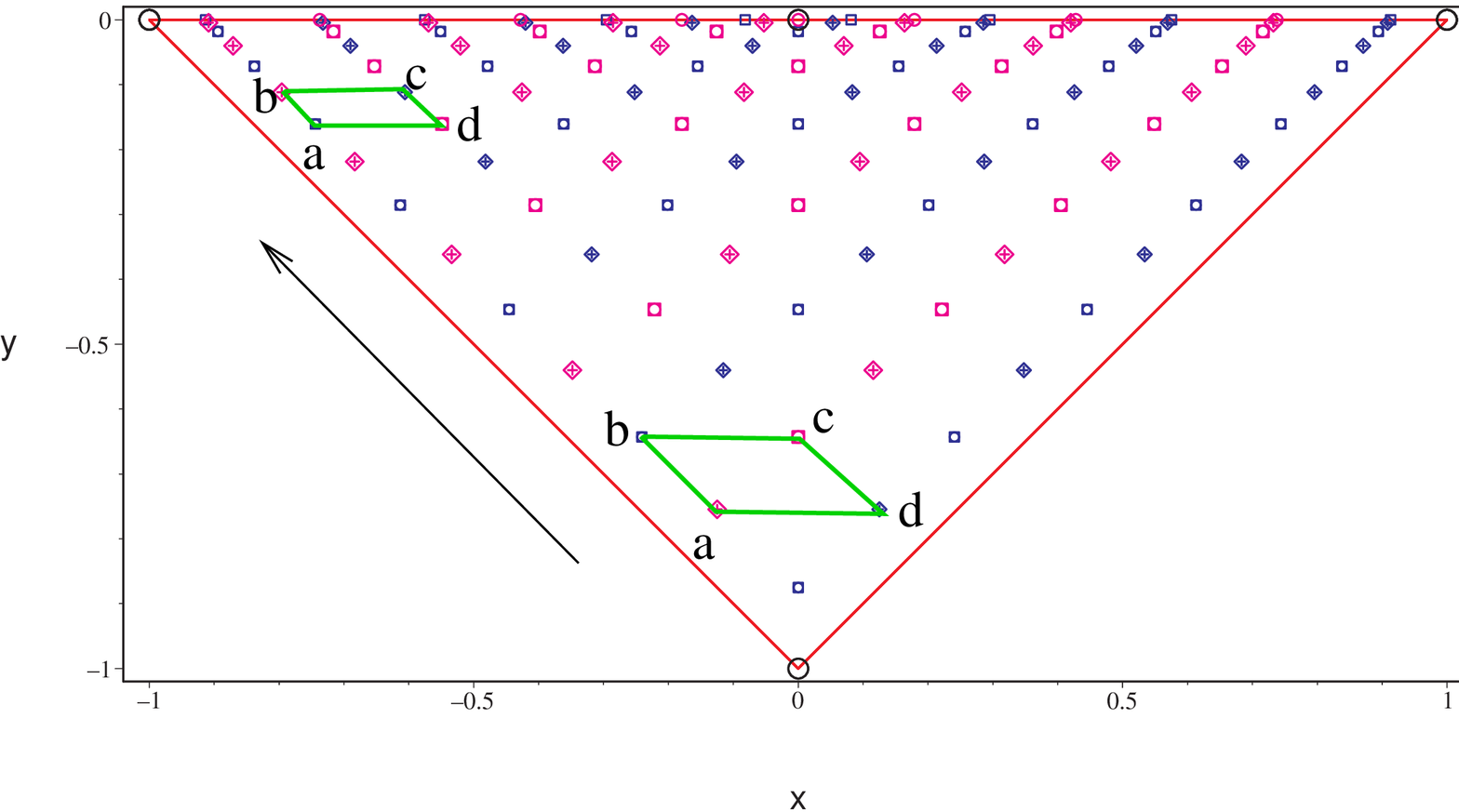}
\end{minipage}\hfil\quad\hfil
\begin{minipage}{75mm}
\includegraphics[width=7cm]{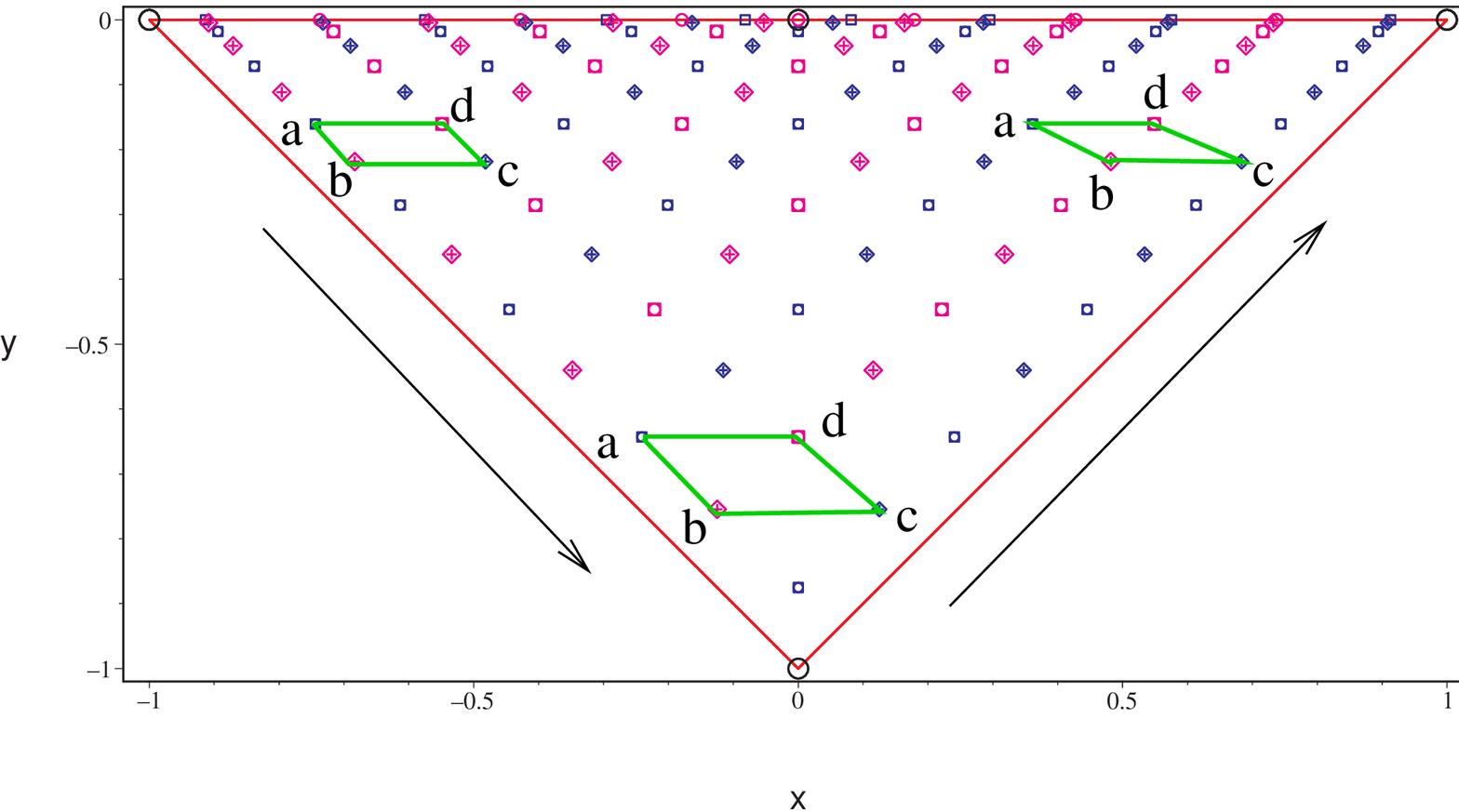}
\end{minipage}

\vspace{2mm}

\centerline{\includegraphics[width=7cm]{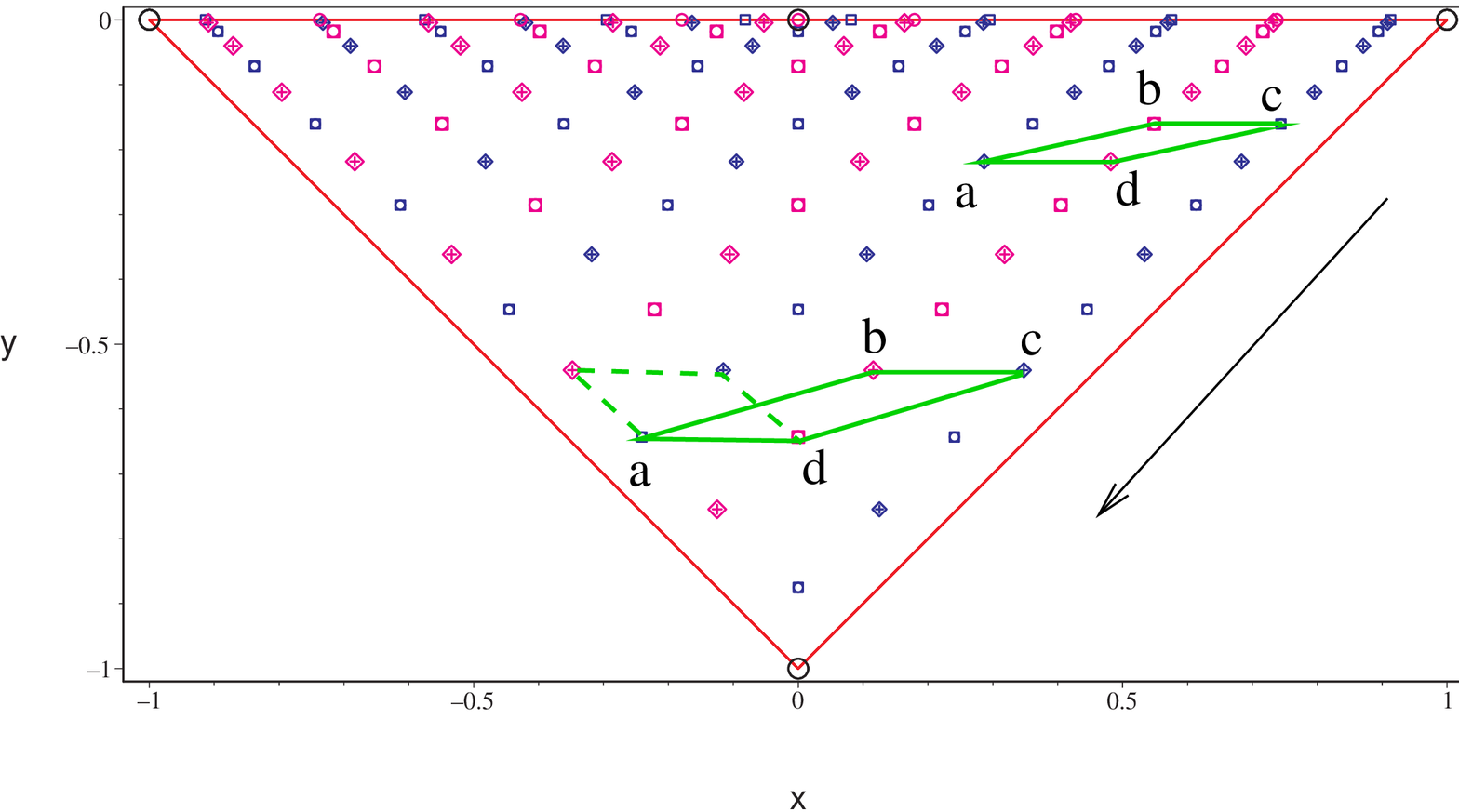}}
\caption{`Parallel transport' of an elementary quantum cell along
the closed path surrounding central singularity for the special limiting
case of Manakov top \protect{(\ref{eq_XY})} with $a=2$, $b=1$.}
\label{Cell_ev_a2_b1}
\end{figure}

The symmetry of the $a=2$, $b=1$ case is higher than the symmetry of a
generic Manakov top problem. Thus we can still use eight irreducible
representations of the initially chosen symmetry group to label
the common eigenvalues even when they belong to doubly degenerate
representations of higher symmetry group.  We
split eight irreducible representations
of the initial symmetry group into two dif\/ferent groups
and associate each group with its own leaf on the EM diagram.
In such a case we start at Fig.~\ref{Cell_ev_a2_b1} (upper sub-f\/igure) with an
elementary cell formed by four dif\/ferent representations and
associated with one leaf. We move the cell towards $Y=0$ stratum and cross it
(using $Y^\prime$ unfolded coordinates) changing at the same time the
irreducible representations associated with the vertices of the cell.
Further evolution (Fig.~\ref{Cell_ev_a2_b1}, middle) is done on the second
leaf. Then passing again through the $Y=0$ stratum we return back
to the f\/irst leaf and can compare f\/inal cell with the initial one.

\begin{figure}[t]
\centerline{\includegraphics[width=6.5cm]{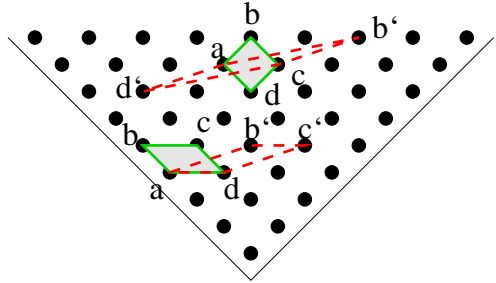}}
\caption{Transformation of the cell after parallel transport along
closed path represented
in Fig.~\protect{\ref{Path}}. Two dif\/ferent choices of
an elementary cell are shown.
Initial cells are shown by solid line. Final cells are shown by
dash line.}
\label{Monod}
\end{figure}

Naturally, the transformation of the cell along the closed path
depends on the choice of the basis. Two alternative choices of the
lattice basis are used in Fig.~\ref{Monod} to illustrate the
monodromy transformation. It is well known that the matrix representation of
the monodromy transformation depends on the lattice basis and  is def\/ined up to
similarity transformation with a~matrix from $SL(2,Z)$
corresponding to basis transformation of regular lattice. For two examples
shown in Fig.~\ref{Monod}
the matrix of the monodromy transformation can be
easily written in an algebraic form. For one case (lower in Fig.~\ref{Monod})
we have
\begin{gather*}
ad \rightarrow ad; \qquad  ab \rightarrow ab + 2ad ,
\end{gather*}
and the corresponding matrix is
$\left(\begin{matrix} 1 & 0 \\ 2 & 1 \end{matrix} \right) $.
Alternative choice of the basis shown on the same f\/igure
gives
\begin{gather*}
ad \rightarrow -2ab - ad; \qquad  ab \rightarrow 3ab + 2ad ,
\end{gather*}
and the corresponding matrix is
$\left(\begin{matrix} 3 & 2 \\ -2 & -1 \end{matrix} \right) $.

%%%%%%%%%%%%%%%%%%%%%%%%%%%%%%%%%%%%%%%%%%%%%%%%%%%%%%%%%%%%%%%%%%%
%%%%%%%%%%%%%%%%%%%%%%%%%%%%%%%%%%%%%%%%%%%%%%%%%%%%%%%%%%%%%%%%%%%%%

%%%%%%%%%%%%%%%%%%%%%%%%%%%%%%%%%%%%%%%%%%%%%%%%%%%%%%%%%%%%%%%%%%%%%%
\subsection{Evolution of quantum cell for generic Manakov top problem}
For the generic Manakov top problem in order to realize the evolution
of the quantum cell along the closed path shown in Fig.~\ref{PathG}
we need to study crossing two dif\/ferent singular strata.

Let us start with crossing $DF$ (or $AF$) stratum. We make the choice of an
elementary quantum cells in region I as formed by four eigenvalues of
dif\/ferent symmetry $(A_s,A_a,B_{1s}, B_{1a})$, associated with leaf 1.
Four other symmetry types $(B_{2s}, B_{2a}, B_{3s}, B_{3a})$ are
associated with another leaf.

The splitting of eight dif\/ferent representations into two groups of
four is unique because we impose the requirement that after the
evolution of all vertices of the cell into region~III or region~IV
the elementary cell should remain elementary, i.e.\ vertices
should not belong, for example, to the pair of degenerate eigenvalues.
The evolution of eigenvalues of each symmetry type is realized
using the correspondence between the joint spectrum lattice
formed by states of one
symmetry type  and the part of the regular square lattice
having the form of an equilateral rectangular triangle. Such
correspondence is global and it enables us to go through the
singular stratum $DF$ or $AF$
(point $b$ or $e$ in Fig. \ref{PathG}).

\begin{figure}[t]
\centerline{\includegraphics[width=6cm]{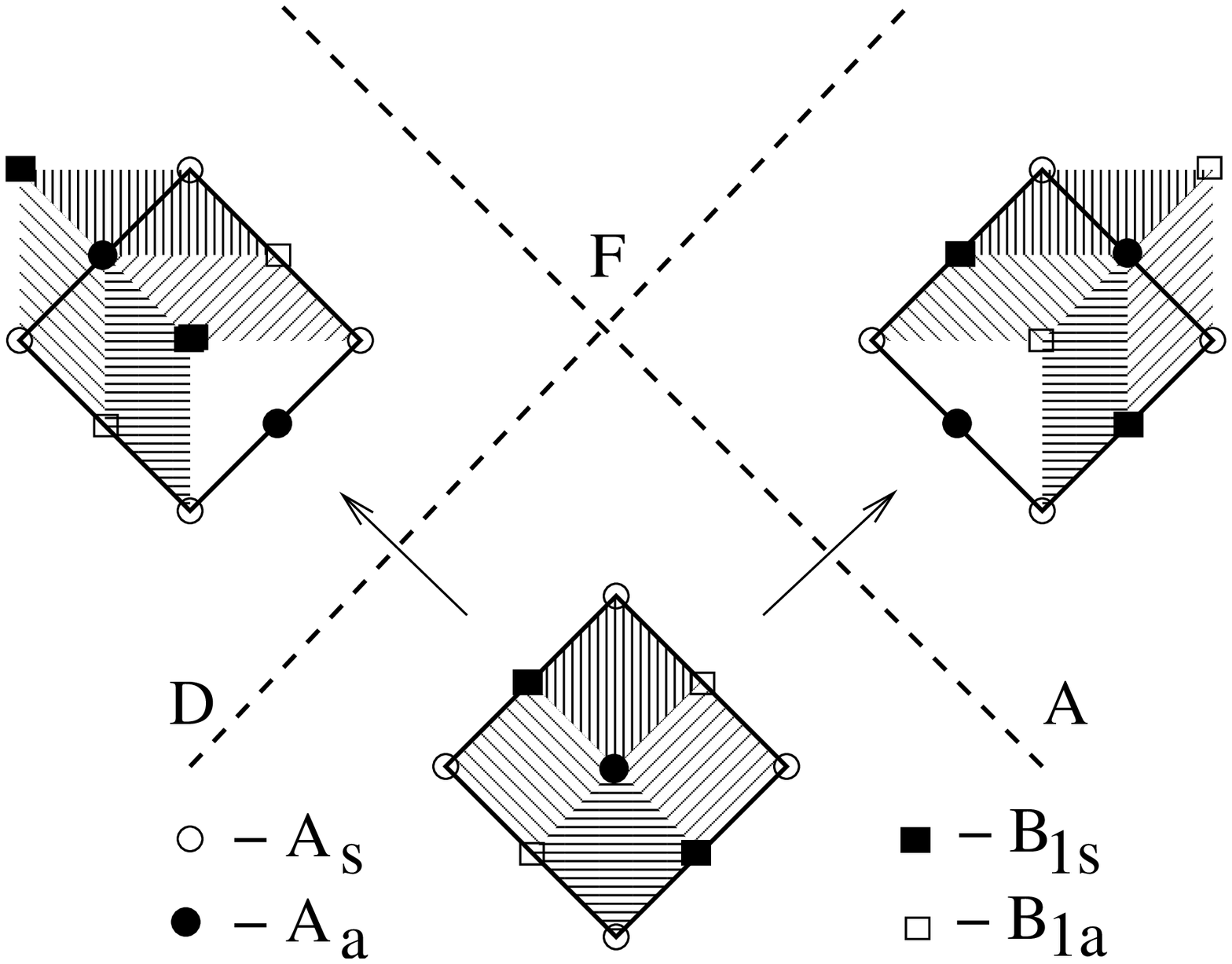} \hskip1cm
\includegraphics[width=6cm]{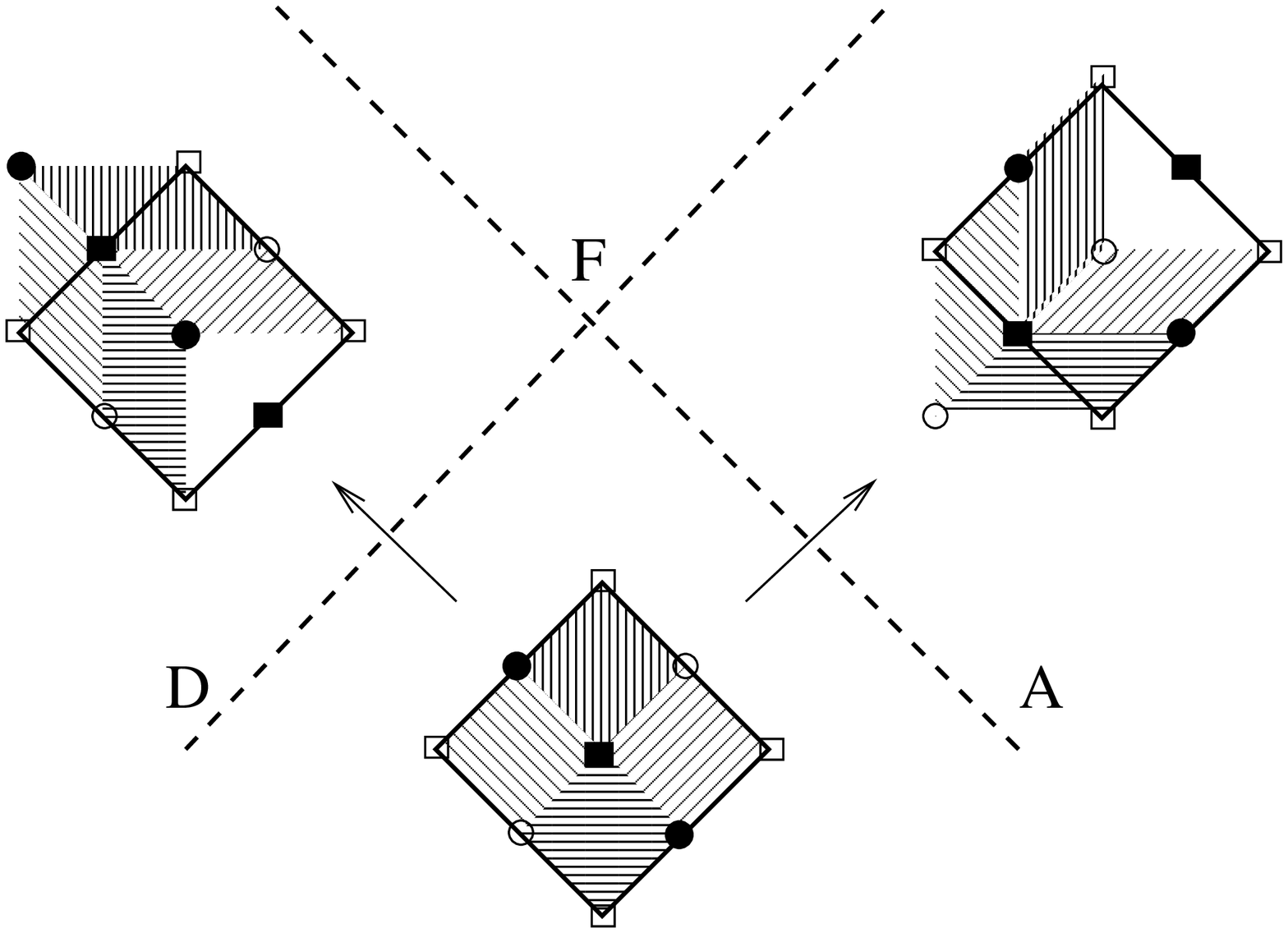}}
\caption{Transformation of quantum cell under crossing singular strata $DF$
and $AF$ (see Fig.~\protect{\ref{Path})}. Elementary cells are
shown by dif\/ferent hatching. Two alternative choices of big `quadruple'
cell are shown. Left: Cell is formed by $A_s$ vertices. Right:
Cell is formed by $B_{1a}$ vertices.}
\label{ASleft}
\end{figure}

Fig.~\ref{ASleft} shows that dif\/ferent elementary cells transforms
after crossing singular stratum in dif\/ferent way. Saying in another way,
this singular stratum is not passable by an elementary cell.
Nevertheless, one can choose bigger cells which pass through singular
stratum unambiguously. The situation here is similar to
the fractional $1/2$ monodromy, where elementary cell cannot pass but the
double cell passes. At the same time,  the case of Manakov top is slightly
dif\/ferent. In order to pass through $DF$ stratum the cell should be
doubled in the direction parallel to~$DF$, whereas to pass through
$AF$ stratum the cell should be doubled in the direction parallel
to~$AF$ which is orthogonal to~$DF$ from the point of view of regular
lattice in the region I. The conclusion: In order to pass through
both $DF$ and $AF$ the cell should be quadruple in such a way that
all its vertices correspond to the same irreducible representation.
Fig.~\ref{ASleft} shows evolution of two such quadruple cells
(with vertices of $A_s$ and of $B_{1a}$ symmetry respectively).
Similar modif\/ications  take place for all other cells
which can be def\/ined on both leafs in the region I.

%%%%%%%%%%%%%%%%%%%%%%%%%%%%%%%%%%%%%%%%%%%%%%%%%%%%%%%%%%%%%%%%%%%
\begin{figure}[t]
\centerline{\includegraphics[width=8cm]{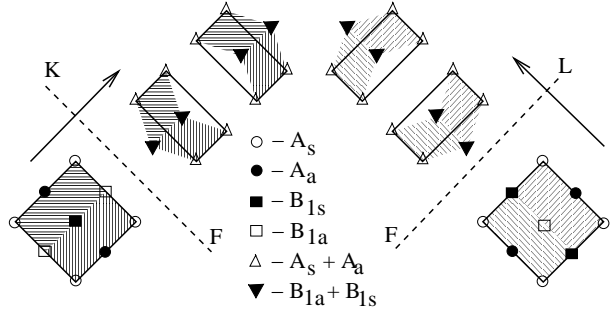}}
\caption{Transformation of quantum cells under crossing singular strata $DF$
and $AF$ (see Fig.~\protect{\ref{Path})}. Minimal (double elementary)
cells are shown by dif\/ferent hatching. Big `quadruple'
cell consists of two minimal cells.
It splits into two cells which belong to dif\/ferent leafs in region II.}
\label{Bid2}
\end{figure}
%%%%%%%%%%%%%%%%%%%%%%%%%%%%%%%%%%%%%%%%%%%%%%%%%%%%%%%%%%%%%%%%%%%%%

The situation with crossing singular strata $KF$ and $FL$ is quite
dif\/ferent from crossing $DF$ or $FA$ strata. Entering region II
is associated with splitting of one connected component
of the classical f\/ibration into two. The associated transformation
of quantum cell is the splitting of one cell into two. The natural
physical requirement imposed on such transformation is the
conservation of the reduced volume \cite{AnnPhys}, i.e. the
volume of the cell in the local action variables. This means that
in order to pass through the $KF$ line from region III to
region II, for example, the initial cell should be at least double.
After crossing $KF$ line it splits in this case into two single cells
associated with two dif\/ferent leafs in region II. These two dif\/ferent
cells can be moved through region II along two-component path
represented in Fig.~\ref{PathG}.

Fig.~\ref{Bid2} shows transformation of quantum cells when
they cross the singular stratum~$KF$ and~$FL$.
The f\/irst important observation is that two double cells
(formed each by two elementary cells in region III and shown by
dif\/ferent hatching) leads to dif\/ferent
pairs of single cells in region II. This means that the minimal cell
should be doubled once more in order to def\/ine an unambiguous
transformation of the quantum cell after crossing singular $KF$
stratum. The quadruple cell chosen in regions III as having
vertices of the same symmetry splits into two double cells in
region II in a unique way.
The situation is completely similar with crossing
$FL$ stratum.

Reverse transformation from region II to region III or IV through
singular $KF$ or $FL$ strata leads to fusion of two double cells
into one quadruple cell.

%%%%%%%%%%%%%%%%%%%%%%%%%%%%%%%%%%%%%%%%%%%%%%%%%%%%%%%%%%%%%%%%%%%
\begin{figure}[t]
\centerline{\includegraphics[width=7cm]{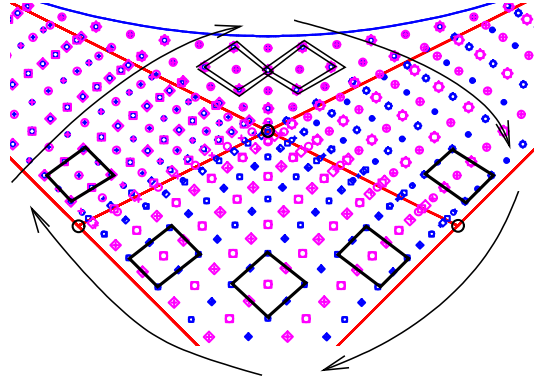}}
\caption{Transformation of quantum cell along a closed path
 (see Fig.~\protect{\ref{PathG})}.
Quadruple cell chosen in region I remains quadruple in region III,
splits into two double cells entering into region II, fuses
again into quadruple cell returning back to region IV and to
region I.}
\label{multi}
\end{figure}
%%%%%%%%%%%%%%%%%%%%%%%%%%%%%%%%%%%%%%%%%%%%%%%%%%%%%%%%%%%%%%%%%%%%%

Now, we can realize the evolution of the quantum cell along the closed
path represented in Fig.~\ref{PathG}. We present in Fig.~\ref{multi} such evolution using the essential part of the joint spectrum
of the
Manakov problem for $a=4$, $b=3$, $S=T=15$ (compare with total joint
spectrum shown in Fig.~\ref{fig_EMtot15}). It should be
noted that we need to follow the evolution of two initial
cells. In the region~I one of these cells belongs to leaf I$_1$,
while another belongs to leaf I$_2$. But as soon as evolution of both
cells is completely similar except for the fact that the cells follow
dif\/ferent leafs, we can discuss only the case of the cell located on the
leaf I$_1$. The initial cell consists of four elementary cells in
the region~I. It has all its vertices labeled by the same
irreducible representation and it is associated with the leaf I$_1$.
We follow further the path represented in Fig.~\ref{PathG}
by solid line. The cell can cross the singular stratum $DF$ and to
move further through leaf III$_1$ till singular stratum $KF$.

When crossing $KF$ stratum the quadruple cell transforms into
two double cells located at two dif\/ferent leafs in region II.
We denote these leafs as II$_{1a}$ and II$_{1b}$ and remind that
there are four dif\/ferent leafs in the region II. The notation we use
is based on the fact that the leaf III$_i$, $(i=1,2)$ splits under crossing
$KF$ into two leafs II$_{ia}$ and II$_{ib}$. In a similar way
the leaf IV$_i$ splits
under crossing $FL$ into two leafs II$_{ia}$ and II$_{ib}$.

Thus in the region II instead of one quadruple cell we have two
double cells which are shown superimposed in Fig.~\ref{multi}.
These two double cells can be moved through the region II with each
cell following its proper leaf. When crossing $FL$ singular stratum
the two cells fuse together and form one quadruple cell located at
the leaf IV$_1$. This cell moves further towards the singular
stratum $FA$, crosses it and returns back on the leaf I$_1$ to the
initial position.

Exactly the same transformation takes place for the initial
quadruple cell chosen in region~I at leaf I$_2$.  In both cases
the transformation between initial and f\/inal cell is trivial, i.e.
the monodromy matrix is identity. The non-triviality of such
transformation is due to the fact that only quadruple cells are
passable and the path along which the cell is propagated has two
branching points where the path bifurcates into two-component path and fuse
from two-component path back into one-component path.

It is quite interesting to compare the present situation with
the analysis of the $1:2$  \cite{SanColin}
and $1:(-2)$ \cite{AnnHP} resonant nonlinear oscillators.
Both these problems have one-dimensional
singular stratum in the image of the momentum map which is
formed by points with inverse image being `curled torus'
\cite{SanColin,AnnHP}. This stratum is not
passable in quantum version by an `elementary cell' but
it is passable by `double cell'.
$1:2$ resonance problem has no nontrivial monodromy because there is no
non-contractible circular paths due to the fact that the line
of critical values ends at the boundary and the end point cannot
be encircled. In contrast the $1:(-2)$  problem~\cite{AnnHP} has singular
stratum with the end point and the nontrivial fractional monodromy
could be introduced with that example. Probably the present
discussion of the quantum Manakov top will stimulate looking for
further examples of integrable systems with still less trivial
but generic behavior in classical and quantum systems.

\section{Conclusions}
The analysis of the possible propagation of the quantum cell through
the joint quantum spectrum of two commuting observables for the
Manakov top model is studied
in this paper for the f\/irst time. The presentation of the material
here is done on completely heuristic physical ground. Nevertheless,
the authors hope that our result about quantum monodromy
will f\/ind more serious description in classical as well as in
quantum mechanics. We believe that further analysis will lead
to the formulation of new important qualitative features of
classical and quantum problems and allow to make further important
steps in formulating general qualitative theory of highly
excited quantum systems which is the ultimate goal of the authors.

\appendix

\section{Symmetry group action on the phase space}
The f\/irst step in the qualitative analysis of any given
model problem is the analysis of the symmetry group action
on the dynamical variables. Below we follow general ideas of the
group theoretical and topological analysis of molecular models
outlined in \cite{Michel1,Michel2}.

Fixing Casimirs $S=T>0$ we obtain direct product
of two $S_2$ spheres as a phase space of the  classical
Manakov top.

\begin{figure}[t]
\centerline{\includegraphics[height=50mm]{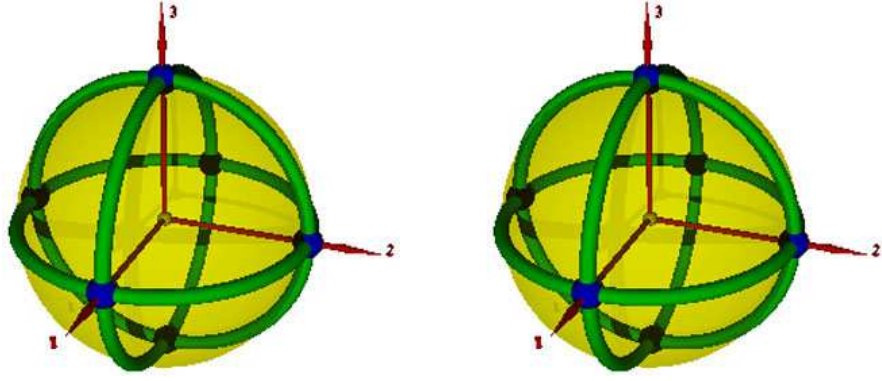}\quad\qquad
%\subfigure
\includegraphics[height=50mm]{Zhilinskii-fig14-new}}
\caption{Diagonal action of $D_{2h}$ group on $S$ and $T$ spheres.}
\label{fig_Stratification_S2}
\end{figure}

Now we want to f\/ind the stratif\/ication of this space under the
symmetry
group action. For model in question the symmetry group $G$
consists of two subgroups. One, which we denote~$D_{2h}$
using standard Sch\"onf\/lies notation, acts in
natural diagonal way on two $S$ and $T$ spheres. Another subgroup
is the permutation of appropriate points on $S$ and $T$ spheres,
$P_{st}$, which acts as $s_i \leftrightarrow t_i$. The total group
is  $D_{2h}\wedge P_{st}$. The diagonal action of
$D_{2h}$ on two  spheres~$S$ and~$T$ is shown on
Fig.~\ref{fig_Stratification_S2}. On every sphere we have
three one dimensional invariant manifolds (bold circles)
with $C_s^{(ij)}$ stabilizers ($C_s^{(ij)}$ is an order two group
generated by the ref\/lection in the plane passing through axes $i$ and $j$).
Three pairs of diametrically
opposite points of intersections of circles form three zero dimensional
strata with $C_{2v}^{(j)}$ stabilizers ($C_{2v}^{(j)}$ is an order
four group generated by two ref\/lections in planes $ij$ and $jk$).
All the remaining points on the
spheres have a trivial stabilizer 1 and so belong to principal type of
orbits (orbits that consist of regular points). The
isolated points of the group action, i.e.\  points with
local symmetry (stabilizer) dif\/ferent from local symmetries of all
other neighboring points, form  critical orbits.
 By the theorem of Michel~\cite{Michel,Michel1} the
gradient of every $G$-invariant function vanishes on  critical
orbits. As a consequence, certain stationary points of invariant
functions can be found using only symmetry group action rather than
the concrete form of functions.

To determine the nontrivial invariant subspaces of $D_{2h}$
group in full four dimensional space we need to look for products of
subspaces on $S$ and $T$ spheres which have
nontrivial intersections of their stabilizers. For example
let's take the point with local symmetry  $C_{2v}^{(1)}$ on
$S$ sphere, then on $T$ sphere we must take the point with the
same stabilizer in order to obtain an orbit of critical points in four
dimensional space. Dif\/ferent combinations of points  on $S$ and $T$
spheres  give us a family of two critical orbits (each formed of two points)
in the full space. All resulting
zero dimensional strata of the Manakov top phase space are
listed in Table~\ref{tab_IsolPoints}. Points of the same
orbit are denoted by one letter. They are further distinguished
by indices.

\begin{table}[t]
\begin{center}\tabcolsep=2.5ex\renewcommand{\arraystretch}{0.9}
\caption{Isolated points and their stabilizers.}
\vspace{1mm}
\begin{tabular}{|c|c|c|c|c|c|c|l|}
\hline
Point &$s_1$ &$s_2$ &$s_3$ &$t_1$ &$t_2$ &$t_3$ &Stabilizer \tsep{2pt}\bsep{1pt}
\\
\hline
$B_2$ &$\phantom{-}1$ &$\phantom{-}$0 &$\phantom{-}$0 &$\phantom{-}1$
&$\phantom{-}$0 &$\phantom{-}$0 &$C_{2v}^{(1)}\wedge P_{st}$ \tsep{6pt}
\\
$B_1$ &$-1$ &$\phantom{-}$0 &$\phantom{-}$0 &$-1$ &$\phantom{-}$0 &$\phantom{-}$0 &\\
$C_2$ &$\phantom{-}1$ &$\phantom{-}$0 &$\phantom{-}$0 &$-1$ &$\phantom{-}$0 &$\phantom{-}$0 &$C_{2v}^{(1)}\wedge P_{st}\sigma^{23} $
\\
$C_1$ &$-1$ &$\phantom{-}$0 &$\phantom{-}$0 &$\phantom{-}1$ &$\phantom{-}$0 &$\phantom{-}$0 &\\
$E_1$ &$\phantom{-}$0 &$\phantom{-}1$ &$\phantom{-}$0 &$\phantom{-}$0 &$\phantom{-}1$ &$\phantom{-}$0 &$C_{2v}^{(2)}\wedge P_{st}$
\\
$E_2$ &$\phantom{-}$0 &$-1$ &$\phantom{-}$0 &$\phantom{-}$0 &$-1$ &$\phantom{-}$0 &\\
$F_1$&$\phantom{-}$0 &$\phantom{-}1$ &$\phantom{-}$0 &$\phantom{-}$0 &$-1$ &$\phantom{-}$0 &$C_{2v}^{(2)}\wedge P_{st}\sigma^{13}$
\\
$F_2$ &$\phantom{-}$0 &$-1$ &$\phantom{-}$0 &$\phantom{-}$0 &$\phantom{-}1$ &$\phantom{-}$0 &\\
$A_1$&$\phantom{-}$0 &$\phantom{-}$0 &$\phantom{-}1$ &$\phantom{-}$0 &$\phantom{-}$0 &$\phantom{-}1$ &$C_{2v}^{(3)}\wedge P_{st}$
\\
$A_2$ &$\phantom{-}$0 &$\phantom{-}$0 &$-1$ &$\phantom{-}$0 &$\phantom{-}$0 &$-1$&\\
$D_1$&$\phantom{-}$0 &$\phantom{-}$0 &$\phantom{-}1$ &$\phantom{-}$0 &$\phantom{-}$0 &$-1$ &$C_{2v}^{(3)}\wedge P_{st}\sigma^{12}$
\\
$D_2$ &$\phantom{-}$0 &$\phantom{-}$0 &$-1$ &$\phantom{-}$0 &$\phantom{-}$0 &$\phantom{-}1$&
\\
\hline
\end{tabular}
       \label{tab_IsolPoints}
    \end{center}
\end{table}

Taking now invariant circles on $S$ and $T$ spheres with the
same stabilizer we can form in the 4-dimensional space their
products. This gives invariant tori
$T_i(s_1,s_2,s_3,t_1,t_2,t_3)$\footnote{This is two
dimensional tori $T_i^2$, but we will omit the index $2$ and
denote them as $T_i$.} in $S_s^2 \otimes S_t^2$ (their
stabilizers are given in the right column):
\begin{gather*}
T_1(0,s_2,s_3,0,t_2,t_3),\qquad G_{T_1}=\{E,\sigma^{23}\},\nonumber
\\
T_2(s_1,0,s_3,t_1,0,t_3),\qquad G_{T_2}=\{E,\sigma^{13}\},\nonumber
\\
T_3(s_1,s_2,0,t_1,t_2,0),\qquad G_{T_3}=\{E,\sigma^{12}\}.
%\label{eq_ToriiStbilizers}
\end{gather*}

As soon as each  basic circle of each torus has four points of
higher symmetry each torus itself  contains eight
isolated (by local symmetry) points. All twelve isolated points
are listed in Table~\ref{tab_IsolPoints}. The number of
$G$-invariant tori is three and on each of them there are
eight points whereas in Table \ref{tab_IsolPoints} we have only
twelve points. This means that every point belongs to
two tori, or, in other words, every isolated
point on a torus is a common  point with another torus.

Finally, acting by $D_{2h}$ group on Manakov top phase space
we obtain three invariant subspaces. They are tori
which have  isolated  points on them in  such a way that
each torus has four common points
with each of two other tori.

\looseness=1 To complete the analysis of the symmetry group action
on the classical phase space we need to consider the action
of the permutations of  $s$ and $t$ components on the
full space and on the $G$-in\-variant tori in particular. In full
space the action of operations $P_{st} g$ ($g\in D_{2h}$) gives
eight invariant subspaces $S_i(s_1,s_2,s_3,t_1,t_2,t_3)$
(for further convenience we  call them $G$-invariant spheres):
%%%%%%%%%%%%%%%%%%%%%%%%%%%%%%%%%%%%%%%%%%%%%%%%%%%%%%%%%%%
\begin{alignat}{4}
& S_1(\alpha,\beta,\gamma,\alpha,\beta,\gamma),\qquad && G_{S_1}=\{E,P_{st}\},\qquad && {\rm dynamically\; invariant},& \nonumber \\
& S_2(\alpha,\beta,\gamma,\alpha,\beta,-\gamma),\qquad && G_{S_2}=\{E,P_{st}\sigma^{12}\},&&&\nonumber \\
& S_3(\alpha,\beta,\gamma,\alpha,-\beta,\gamma),\qquad && G_{S_3}=\{E,P_{st}\sigma^{13}\},&&& \nonumber \\
& S_4(\alpha,\beta,\gamma,-\alpha,\beta,\gamma),\qquad && G_{S_4}=\{E,P_{st}\sigma^{23}\},&&&\nonumber \\
& S_5(\alpha,\beta,\gamma,\alpha,-\beta,-\gamma),\qquad && G_{S_5}=\{E,P_{st}C_2^{(1)}\},\qquad && {\rm dynamically\; invariant},&
\nonumber \\
& S_6(\alpha,\beta,\gamma,-\alpha,\beta,-\gamma),\qquad && G_{S_6}=\{E,P_{st}C_2^{(2)}\},\qquad && {\rm dynamically\; invariant},&
\nonumber \\
& S_7(\alpha,\beta,\gamma,-\alpha,-\beta,\gamma),\qquad && G_{S_7}=\{E,P_{st}C_2^{(3)}\},\qquad && {\rm dynamically\; invariant},&
\nonumber \\
& S_8(\alpha,\beta,\gamma,-\alpha,-\beta,-\gamma),\qquad && G_{S_8}=\{E,P_{st}I\}.&&&
\label{eq_SphereStbilizers}
\end{alignat}
%%%%%%%%%%%%%%%%%%%%%%%%%%%%%%%%%%%%%%%%%%%%%%%%%%%%%%%%%%%%%

In order to describe the group action we introduce on each torus
angle variables:
\begin{alignat}{5}
& T_i:\qquad && s_i=0,\qquad && s_j=\cos \phi_s, \qquad && s_k=\sin \phi_s,& \nonumber
\\
&&& t_i=0,\qquad && t_j=\cos \phi_t, \qquad && t_k=\sin \phi_t.&
\label{eq_ToriiCoords}
\end{alignat}
The action of $P_{st} g$  on $\phi_s$ and $\phi_t$ is shown
for $T_2(s_1,s_3,t_1,t_3)$
in Table \ref{tab_PstActionT2}. Fourth column indicates
four dif\/ferent relations between angle coordinates on torus
resulting in points with higher symmetry. This
symmetry group is given in the last column. The indicated
lines pass throw the isolated points on the torus.

By comparing the stabilizers of lines (Table
\ref{tab_PstActionT2}) and that of isolated points
(Table \ref{tab_IsolPoints}) it is easy to conclude
what line contains
what points. Moreover, the intersections of line stabilizers
with stabilizers of $G$-invariant spheres
(\ref{eq_SphereStbilizers}) are also nontrivial. This means
that all points of the line $\phi_s=\phi_t$, for example, at
the same time belong to $S_1$ ($G_{S_1}\cap G_{(\phi_s=\phi_t)}
\neq \{E\}$) and $S_3$  ($G_{S_3}\cap G_{(\phi_s=\phi_t)}
\neq \{E\}$) invariant subspaces, the line $\phi_s=-\phi_t$
belongs to $S_2$ and $S_5$ invariant subspaces and so on.
The construction of tables equivalent to Table
\ref{tab_PstActionT2} for $T_1$ and $T_3$ shows that the line
equations are the same:
\begin{gather}
\label{eq_LinesEquations}
    \phi_s=\phi_t,\qquad \phi_s=-\phi_t,\qquad
    \phi_s=\pi-\phi_t,\qquad \phi_s=\pi+\phi_t,
\end{gather}
but the stabilizers are dif\/ferent.
Subsequently, every line (\ref{eq_LinesEquations}) on a torus
is an intersection line of this torus with two spheres (one of
these spheres is dynamically invariant) and every torus has lines of
intersections with all $G$-invariant spheres.

\begin{table}[t]
\begin{center}\tabcolsep=2.5ex\renewcommand{\arraystretch}{1.15}
\caption{Action of $P_{st}g$, ($g \in D_{2h}$) on $T_2$.}
\vspace{1mm}
\begin{tabular}{|l|c|c|c|l|}
\hline
Operator & \multicolumn{2}{c|}{Action on $\{s_i,t_i\}$} & Line Equation & \ \quad Stabilizer of Line  \tsep{2pt}\bsep{1pt}
\\
\hline
$P_{st}$ &$s_1=t_1$ &$s_3=t_3$ &$\phi_s=\phi_t$ &$\{E,\sigma^{13},P_{st},P_{st}\sigma^{13}\}$\tsep{4pt}
\\
$P_{st}\sigma^{13}$&$s_1=t_1$&$s_3=t_3$&$\phi_s=\phi_t$&
\\
$P_{st}\sigma^{12}$&$s_1=t_1$&$s_3=-t_3$&$\phi_s=-\phi_t$&$\{E,\sigma^{13},P_{st}\sigma^{12},P_{st}C_2^{(1)}\}$
\\
$P_{st}C_2^{(1)}$&$s_1=t_1$&$s_3=-t_3$&$\phi_s=-\phi_t$&
\\
$P_{st}\sigma^{23}$&$s_1=-t_1$&$s_3=t_3$&$\phi_s=\pi-\phi_t$&$\{E,\sigma^{13},P_{st}\sigma^{23},P_{st}C_2^{(3)}\}$
\\
$P_{st}C_2^{(3)}$&$s_1=-t_1$&$s_3=t_3$&$\phi_s=\pi-\phi_t$&
\\
$P_{st}C_2^{(2)}$&$s_1=-t_1$&$s_3=-t_3$&$\phi_s=\pi+\phi_t$&$\{E,\sigma^{13},P_{st}C_2^{(2)},P_{st}I\}$
\\
$P_{st}I$&$s_1=-t_1$&$s_3=-t_3$&$\phi_s=\pi+\phi_t$&
\\
\hline
\end{tabular}
\label{tab_PstActionT2}
    \end{center}
\end{table}

\begin{figure}[t]
\centerline{\includegraphics[width=135mm]{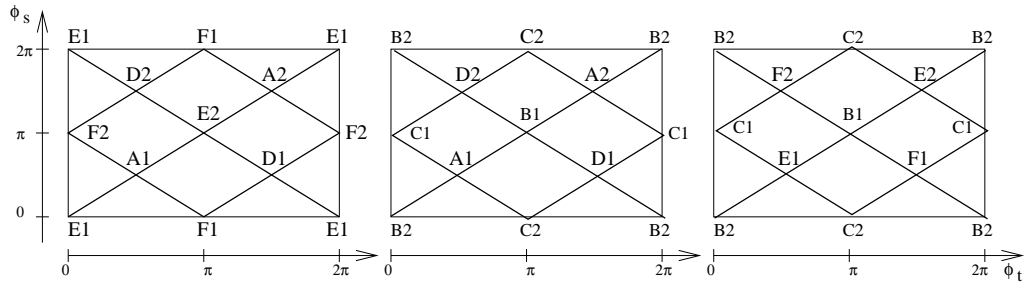}}
  \caption{Stratification of invariant tori.
     Representation of $T_1$, $T_2$ and $T_3$ (from left to right).
Zero-dimensional strata together with their
stabilizers are given in Table~\protect{\ref{tab_IsolPoints}}.
One-dimensional strata are listed in Table~\protect{\ref{tab_PstActionT2}}
for one of the torus, $T_2$. The construction
is similar for two other tori.}
  \label{fig_ToriiMap}
\end{figure}
%%%%%%%%%%%%%%%%%%%%%%%%%%%%%%%%%%%%%%%%%%%%%%%%%%%%%%%%%%%%%

    Thus the stratif\/ication of the Manakov-top phase
space under the action of $D_{2h}\wedge P_{st}$ group
is constructed. In particular, the system of
isolated critical orbits which is due to the symmetry group
action is given. All points forming these orbits are
stationary
points of any invariant function \cite{Michel,Michel1}.
This result is in fact independent on the concrete form
of integrals of motion and relies only on symmetry
arguments. Such preliminary symmetry analysis is quite
important in the qualitative study of molecular models
as it is formulated in our previous works
 \cite{SadZhilin,Michel1,Michel2,KostaSIAM}.

\section{Critical points of energy-momentum map}
In this section we f\/ind critical points of the energy
momentum map def\/ined on the classical phase space of the
Manakov top by two integrals of motion $X$, $Y$ given in
equation~(\ref{eq_XY}). By def\/inition, the point is  critical,
if the  dif\/ferentials of two integrals of motion, $dX$ and $dY$
are linearly dependent,  or equivalently the corresponding
matrix of derivatives has non-maximal rank.
We can greatly simplify the problem
of searching critical points by restricting dif\/ferentials on
the symmetry invariant sub-manifolds.

\begin{figure}[t]
\centerline{\includegraphics[height=40mm]{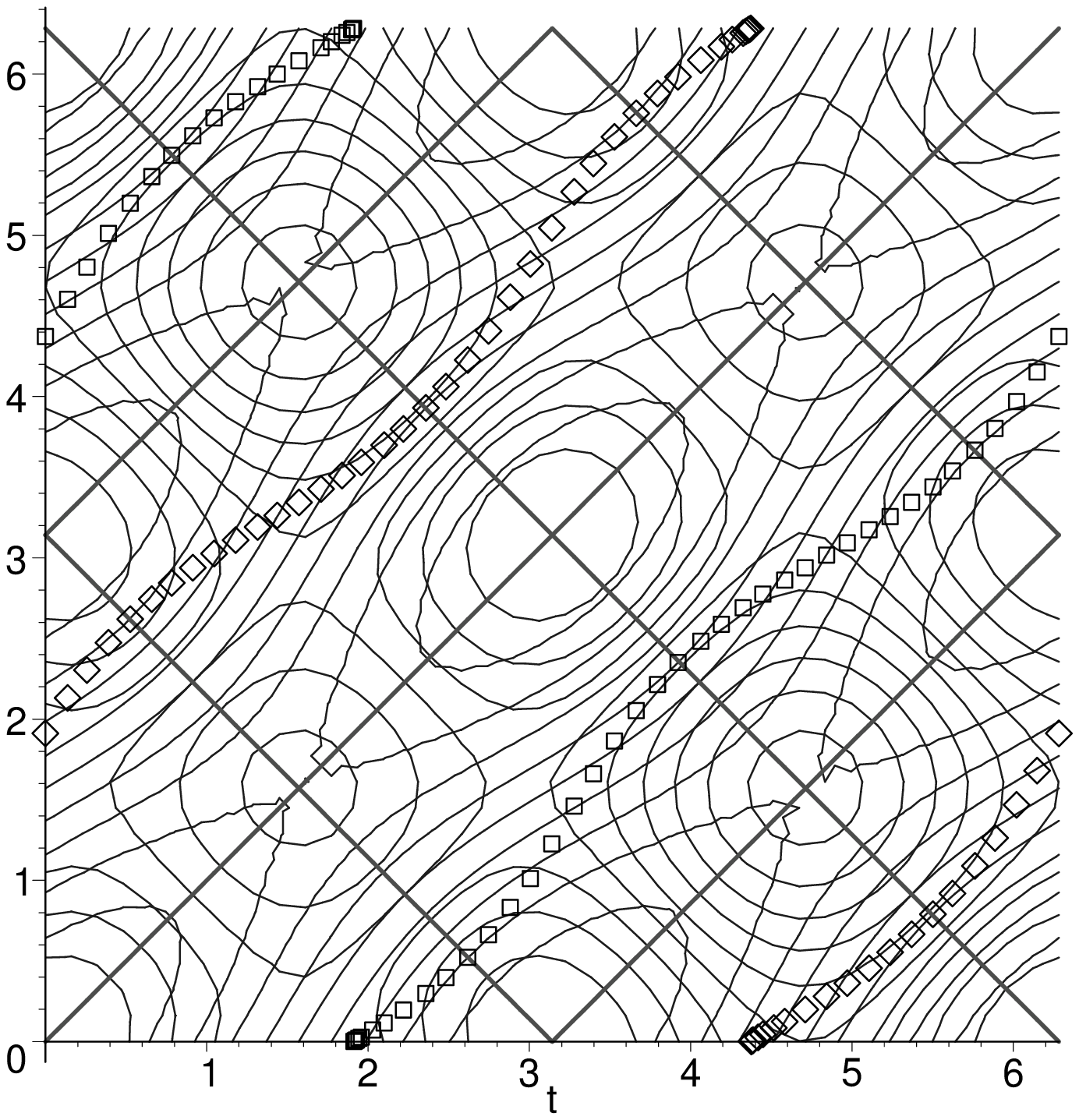}\qquad\quad
\includegraphics[height=40mm]{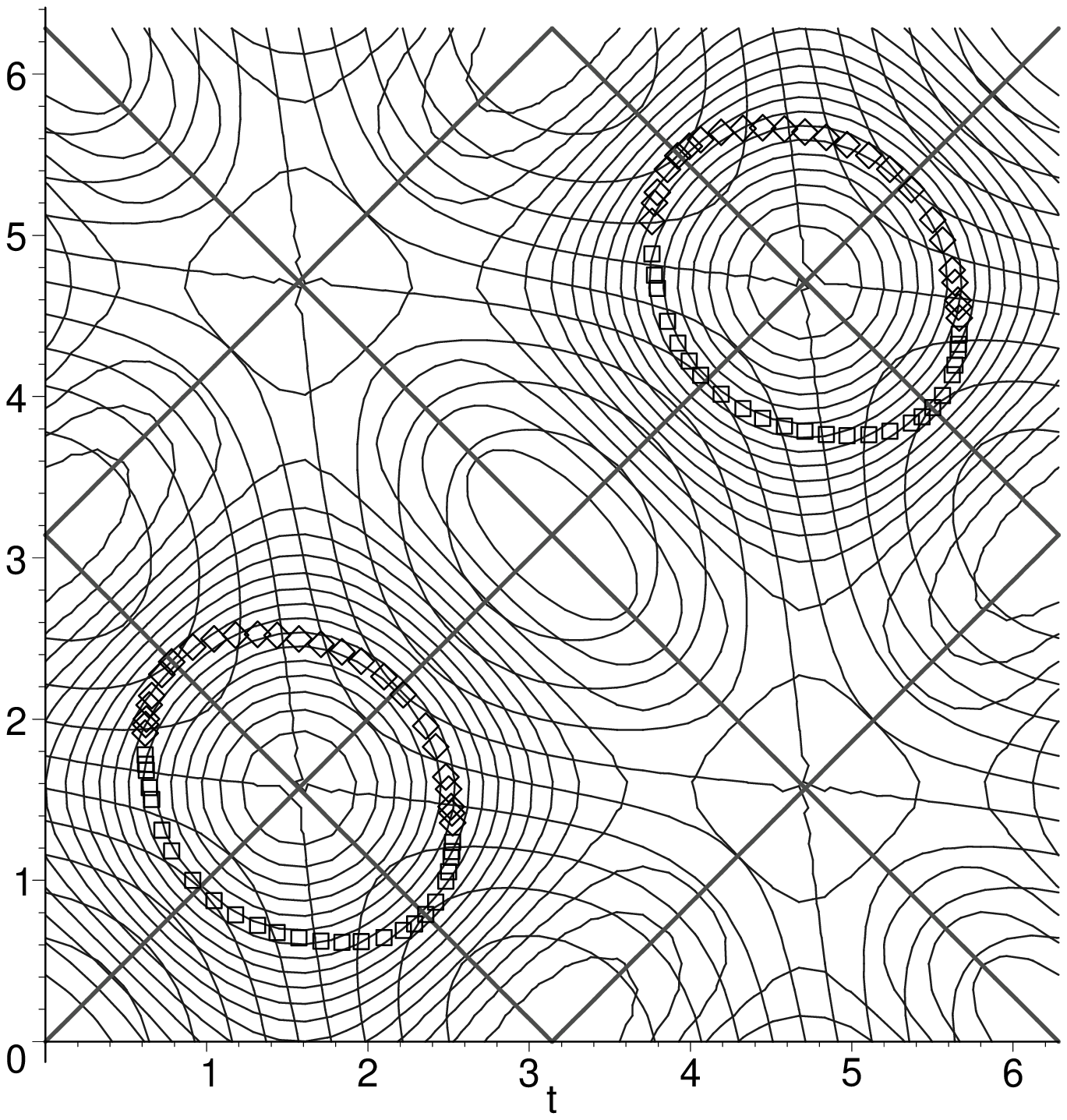}}
\caption{Af\/f\/ine charts of tori 2 and 3  with level lines of $X$ and $Y$ ($a=4$, $b=2$).
Critical curves~(\ref{eq_ParabOnTorii}) are shown by box and diamond points.}
\label{fig_SetkaTora}
\end{figure}

For $G$-invariant tori
with coordinates (\ref{eq_ToriiCoords}) the condition
of non-maximal rank is
\begin{gather}
\label{eq_det}
{\rm det} \left(
  \begin{matrix}
    \dfrac{\partial X}{\partial \phi_s}&\dfrac{\partial X}{\partial \phi_t}\\[2ex]
    \dfrac{\partial Y}{\partial \phi_s}&\dfrac{\partial Y}{\partial \phi_t}
  \end{matrix}
\right)=0.
\end{gather}
The solution of equation~(\ref{eq_det}) has for every torus
($T_1, T_2$, or $T_3$) six roots.
Four of them are, naturally, the lines (\ref{eq_LinesEquations})
\begin{gather*}
%\label{eq_LinesOnTorii}
s=\biggl\{t, -t, \frac1t, -\frac1t\biggr\},
\end{gather*}
which have the same form for $T_1$, $T_2$, $T_3$. These lines are shown in
Fig.~\ref{fig_ToriiMap}. They
are completely def\/ined by the symmetry group action and
do not depend nor on values of
$a$ and $b$ parameters nor  on the choice of invariant functions
def\/ined on $S_s^2\otimes S_t^2$.
Two others roots are curves which depend on the  concrete
form of functions $X$ and $Y$ specif\/ied by parameters $a$, $b$.
Their analytic form varies slightly with invariant torus as follows:
\begin{gather}
T_1: \quad s=\frac{(a-b-1)t\pm\sqrt{a(1-b)(t^4-1)
+2(2b-a(1+b))t^2}}{at^2+b-1}, \nonumber\\
T_2: \quad s=\frac{(a+b-1)t\pm\sqrt{(a-b)(t^4+1)+2(b(a-1)
+a(b-1))t^2}}{-t^2+a-b}, \nonumber\\
T_3: \quad s=\frac{(1-a-b)t\pm\sqrt{(b-a)(t^4+1)+2(b(a-1)
+a(b-1))t^2}}{t^2+a-b}.
\label{eq_ParabOnTorii}
\end{gather}
Here $s=\tan(\phi_s/2)$ and  $t=\tan(\phi_t/2)$.
The form and position of curves (\ref{eq_ParabOnTorii}) on tori $T_i$
depend on $a$ and $b$.
 For  the parameters values $a>b>1$ only two curves from
(\ref{eq_ParabOnTorii}), namely curves def\/ined for $T_2$ and $T_3$
tori, have a real range of values, they  are shown on Fig.~\ref{fig_SetkaTora}.

%%%%%%%%%%%%%%%%%% Setka Tora %%%%%%%%%%%%%%%%%%%%%%%%%%%%%
%%%%%%%%%%%%%%%%%%%%%%%%%%%%%%%%%%%%%%%%%%%%%%%%%%%%%%%%%%%%%
Similar construction  of matrix for $dX$ and $dY$ for
$G$-invariant
spheres gives us the matrix with determinant identically equal to
zero (rank of the matrix is less then two). The rank of this matrix
equals
identically zero  in all points which belong to critical
orbits listed in Table~\ref{tab_IsolPoints}.
 In all other points of the $G$-invariant
spheres the rank of the matrix equals one. This means that there are
no other critical points on spheres except those isolated
points found earlier from the symmetry considerations.

\section{Classical energy momentum map}
The set of two integrals $F=\{X,Y\}$ (\ref{eq_XY}) def\/ines the mapping
$F: S_s^2 \otimes S_t^2 \rightarrow R^2$. Possible values
$f\in R^2$ of the map form the image of the map, or the base of
the corresponding integrable f\/ibration. We distinguish regular
and critical values of the map. The set of critical values
forms what is often called the bifurcation diagram. Below, we
call the set of all, regular and critical values of EM map
an energy-momentum diagram. In this section we discuss the form
of the energy-momentum diagram for one particular choice of parameters
$a$, $b$ and its variation along the modif\/ication of parameters.

Let us f\/irst take parameters $a$, $b$ in (\ref{eq_XY}) as
\begin{equation}
\label{eq: ChoisOfab}
a>b>1,
\end{equation}
and f\/ind images of critical points of rank zero
(Table \ref{tab_IsolPoints}). Substitution of their coordinates
into equation~(\ref{eq_XY}) conf\/irms that every point from one symmetry
group orbit naturally
gives  the same value of functions $X$ and  $Y$, so we have
six critical values of EM map specif\/ied below by their
coordinates in $R^2$ plane of $(X, Y)$ values:
\begin{gather*}
B(1,0),\qquad C(-1,0),\nonumber
\\
E\biggl(\frac{a-b-1}{1-a-b}, \frac{-4ab(1-a)}{1-a-b}\biggr),\qquad
F\biggl(-\frac{a-b-1}{1-a-b}, \frac{4b(1-a)(1-b)}{1-a-b}\biggr),\nonumber
\\[1ex]
A\biggl(\frac{b-a-1}{1-a-b}, \frac{-4ab(1-b)}{1-a-b}\biggr),\qquad
D\biggl(-\frac{b-a-1}{1-a-b}, \frac{4a(1-a)(1-b)}{1-a-b}\biggr).
%\label{eq: PointsImages}
\end{gather*}
Points $B$, $C$ do not depend on $a$, $b$ parameters. For the choice
of parameters as in equation~(\ref{eq: ChoisOfab}), the $x$ component for
all other points always lie in the $[-1,1]$ region and  $y$ component
is  always negative.

\begin{figure}[t]
\centerline{\includegraphics[width=110mm]{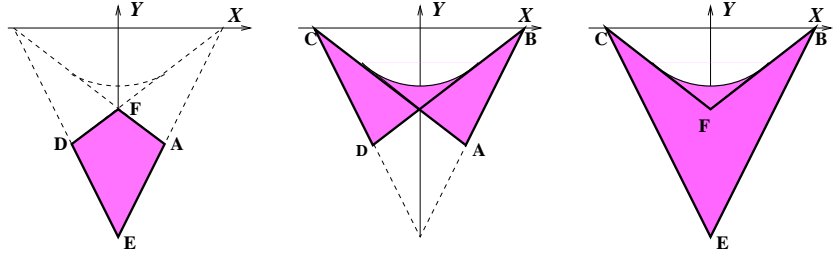}}
  \caption{Images of $T_1$, $T_2$ and $T_3$ tori under the
EM map. Images of zero rank points on tori are denoted
by Latin letters. }
  \label{fig_ToriiImage}
\end{figure}

Next we construct the images under the EM map of
$G$-invariant tori (for $T_i$ we have $s_i=t_i=0$).
In Fig.~\ref{fig_ToriiImage} the colored regions correspond to
images of regular
points on tori, the bold lines  are the images of critical
lines on tori. As it was mentioned earlier, for the given choice of
$a$, $b$ parameters the curves (\ref{eq_ParabOnTorii}) exist
only for two tori. On the EM diagram the image of these curves
is represented as  a part of parabola. We will call
the  neighboring area  having the form of a small curved
triangle the parabola area.
%%%%%%%%%%%%%%%%%% Tori Images  %%%%%%%%%%%%%%%%%%%%%%%%%%%%%
%%%%%%%%%%%%%%%%%%%%%%%%%%%%%%%%%%%%%%%%%%%%%%%%%%%%%%%%%%%%%
Images of lines of critical points on tori (\ref{eq_LinesEquations})
associated with linear dependence between $dX$ and $dY$
can be written explicitly as
\begin{gather*}
Y=2ab(X-1),         \qquad  Y=2a(1-a)(X+1),  \nonumber
\\
Y=2(1-a)(1-b)(X-1), \qquad  Y=2b(1-b)(X+1).
 %\label{eq_EMLinesEq}
 \end{gather*}
\looseness=-1 Explicit form for the boundary of the  parabola region
is simple  only for a certain
choices of~$a$ and~$b$, for example for $a-b=1$ when the
EM diagram
has a symmetric form like that on Fig.~\ref{fig_ToriiImage}.

\begin{figure}[t]
\centerline{\includegraphics[width=150mm]{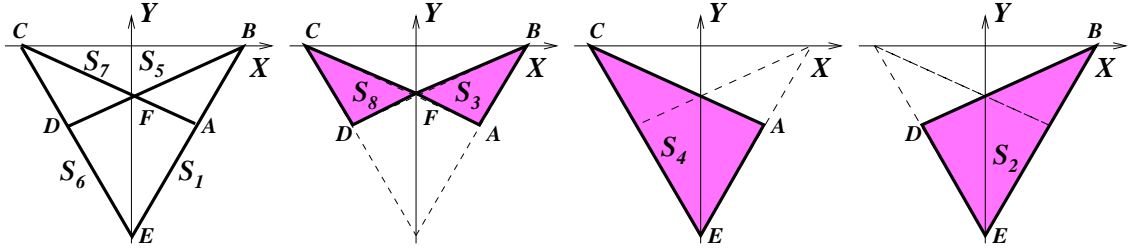}}
  \caption{Images of the $G$-invariant spheres. First f\/igure on the
  left represent four dynamically $G$-invariant spheres
  having 1D-images (straight lines).}
  \label{fig_SphereImage}
\end{figure}

The images of four $G$-invariant spheres  belong to critical
lines (\ref{eq_LinesEquations}), they are shown on
the most left sub-f\/igure of
Fig.~\ref{fig_SphereImage}. Each of these  lines  is
denoted by  $L_i$ with its index corresponding to a
dynamical sphere $S_i$,
$i\in\{1,5,6,7\}$.  All other $S_i$
$i\in\{2,3,4,8\}$ are mapped to 2-D triangular regions in $\cal{EM}$
diagram formed by
regular values and bounded by lines of critical values.
%%%%%%%%%%%%%%%%%% Sphere Images %%%%%%%%%%%%%%%%%%%%%%%%%%%%%
%%%%%%%%%%%%%%%%%%%%%%%%%%%%%%%%%%%%%%%%%%%%%%%%%%%%%%%%%%%%%

So the boundaries  of the image of EM map of Manakov top
for some particular choice of parameters $a$, $b$
are formed by four straight lines and one parabola. There are four
areas: area of parabola (II), two triangles (III, IV) and a
rhomb (I)  (Fig.~\ref{fig_EMgencase}). Two lines always pass
through the f\/ixed point $(-1,0)$ ($L_6$ and $L_7$) and
two other lines always pass through point
$(1,0)$ ($L_1$~and~$L_5$). When
both parameters are of the same sign, $Y$ function is negative
and takes the zero values only in mentioned points.

%%%%%%%%%%%%%%%%%% ab Plane map %%%%%%%%%%%%%%%%%%%%%%%%%%%%%
\begin{figure}[t]
\centerline{\includegraphics[width=85mm]{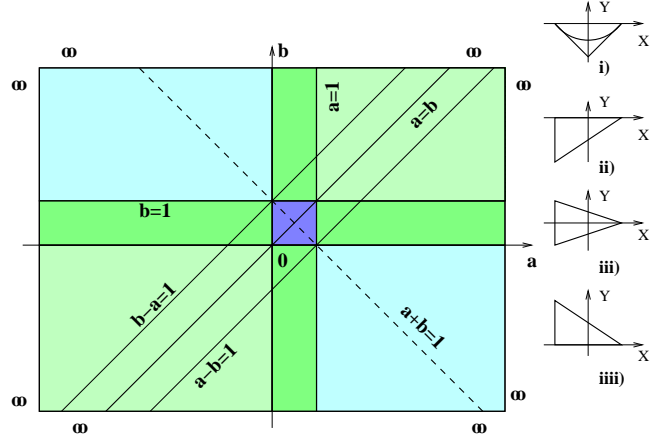}}
  \caption{`Critical' values of $a$ and $b$ parameters (left).
Right f\/igures are the limiting cases correspon\-ding to
$a\rightarrow \infty$ and to : i) $b=a-1$,
ii) $a-1>b\geq 1$, iii) $1>b>0$, iiii) $0\geq b>-a+1$. }
  \label{fig_abPlane}
\end{figure}
%%%%%%%%%%%%%%%%%%%%%%%%%%%%%%%%%%%%%%%%%%%%%%%%%%%%%%%%%%%%%

Now we will summarize brief\/ly the dependence of the
bifurcation diagram on the values of~$a$ and~$b$ parameters.
In the space of~$a$,~$b$ parameters there are regular values
corresponding to qualitatively the same generic diagram
formed by four straight lines and one parabola. Critical values
of~$a$,~$b$ parameters correspond to some degenerate situations
when the image of the EM map qualitatively changes, i.e.\
some regions shrink to zero and some lines coincide.
Moreover, there is some symmetry in the space of $a$, $b$ parameters
which enables one to study only part of the whole plane to
recover all the qualitatively dif\/ferent cases of bifurcation
diagrams. Fig.~\ref{fig_abPlane} which can be named
with some abuse of the language as a `bifurcation diagram
of the Manakov-top-bifurcation-diagram' illustrates the
symmetry in the parameter space.

The critical values of parameters are the
following lines in $R^2$
plane of $(a,b)$ values: $a=0, 1, \pm \infty$; $b=0, 1,
\pm \infty $; $a=b$ and $a+b=1$. The last line (dashed on
Fig.~\ref{fig_abPlane}) is the most degenerate one (all
boundary lines are parallel or coincide: $L_6=L_7$, $L_1=L_5$
and $L_6\| L_1$). Two lines $a=b$ and $a+b=1$ are the symmetry
lines: operations of ref\/lection in the parameter space
$(a,b)\rightarrow (b,a)$, and $(a,b)\rightarrow (-a+1,-b+1)$
do not modify equations def\/ining boundaries of regular
regions on the image of EM map for Manakov top.
The last two lines $b=a\pm 1$
on Fig.~\ref{fig_abPlane} specify the values of parameters
corresponding to the situation when the
bifurcation diagram is  symmetric with respect
to $Y$ axis.

\begin{figure}[t]
\centerline{\includegraphics[angle=270,width=40mm]{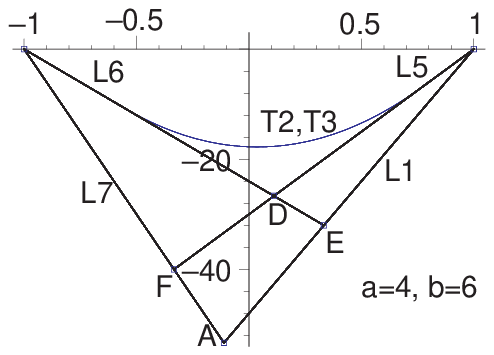}\qquad
  \includegraphics[angle=270,width=40mm]{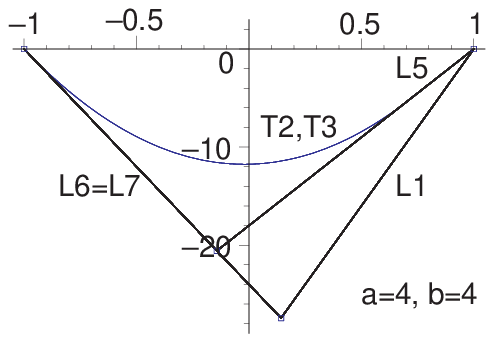}\qquad
  \includegraphics[angle=270,width=40mm]{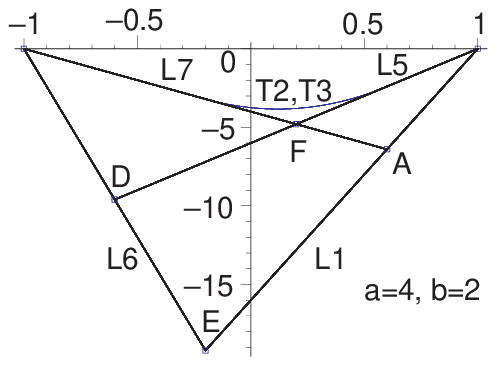}}
  \caption{Bifurcation diagrams for  Manakov top
   corresponding to f\/ixed $a=4$ value and to $b=6, 4, 2$.}
  \label{fig_a4b42}
\end{figure}
%%%%%%%%%%%%%%%%%%%%%%%%%%%%%%%%%%%%%%%%%%%%%%%%%%%%%%%%%%%%%

%%%%%%%%%%%%%%%%%% EM map %%%%%%%%%%%%%
\begin{figure}[t]
\centerline{\includegraphics[width=40mm]{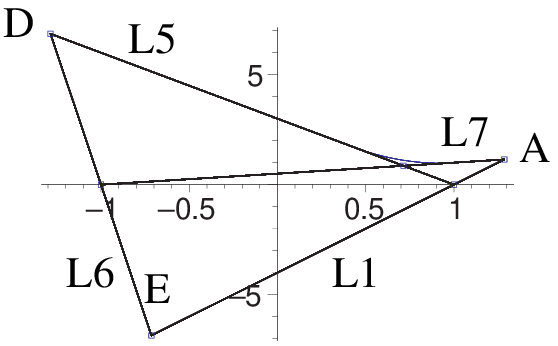}\qquad
  \includegraphics[width=40mm]{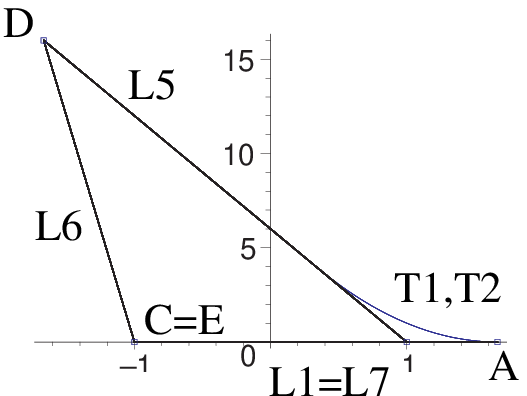}\qquad
  \includegraphics[width=40mm]{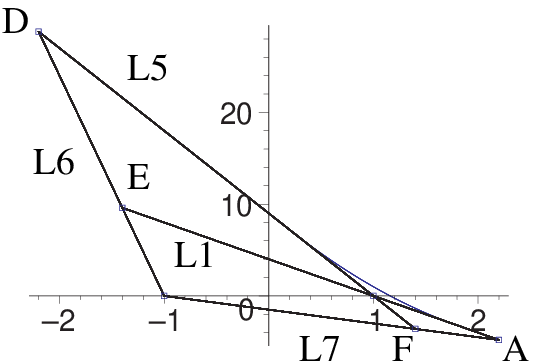}}
  \caption{Bifurcation diagrams for  Manakov top
  corresponding to f\/ixed value of $a$ and to three
  dif\/ferent values of $b=0.5, 0, -0.5$.}
  \label{fig_a4b050-05}
\end{figure}

Some modif\/ications of bifurcation diagram under the
variation of $a,b$ parameters are shown in  Fig. \ref{fig_a4b42}
and in  Fig.~\ref{fig_a4b050-05}.

\begin{figure}[t]
 \centerline{\includegraphics[width=70mm]{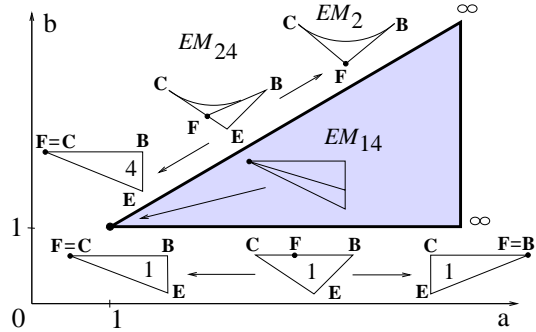}}
\caption{Limiting cases of bifurcation diagram ($a>b>1$).
 Small f\/igures are the images of EM diagram for `critical'
 values of $a$, $b$ parameters and the numbers show what
 regions of the image  do not vanish in that limit.
 The capital letters
 indicate the position of zero rank critical values of
 EM map.}
\label{fig_Rab1}
\end{figure}

\begin{figure}[t]
 \centerline{\includegraphics[width=70mm]{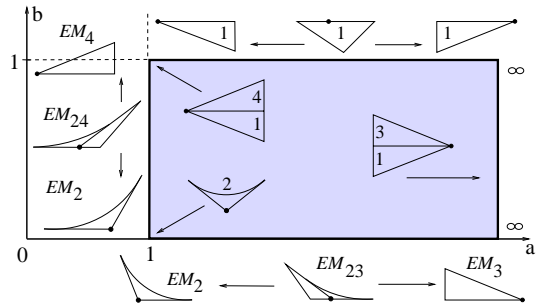}}
\caption{Limiting cases of EM diagram ($a>1$, $1>b>0$).}
\label{fig_Rab2}
\end{figure}

Dif\/ferent limiting cases of the bifurcation diagram
are represented in Figs.~\ref{fig_Rab1} and \ref{fig_Rab2}.

%%%%%%%%%%%%%%%%%%%%%%%%%%% Limiting cases %%%%%%%%%%%%%%%%%%%%%%%
%%%%%%%%%%%%%%%%%%%%%%%%%%%%%%%%%%%%%%%%%%%%%%%%%%%%%%%%%%%%%

It is quite interesting to see the correspondence between images
of EM maps for dif\/ferent limiting cases of Manakov top and
EM diagrams for geodesic f\/low on three-dimensional ellipsoids
with partially coinciding semi-axes \cite{DavDulBol,DavDul}.

\subsection*{Acknowledgments}
This work was stimulated by the initial discussions with Vadim Kuznetsov.
Authors thank Dr. D. Sadovskii for many fruitful discussions and Dr.
C. Davison for discussing with us his recent results submitted
for publication \cite{DavDulBol,DavDul}.

\pdfbookmark[1]{References}{ref}
\LastPageEnding

\end{document}